\documentclass[sigconf]{acmart}
\AtBeginDocument{%
  \providecommand\BibTeX{{%
    \normalfont B\kern-0.5em{\scshape i\kern-0.25em b}\kern-0.8em\TeX}}}

\copyrightyear{2022}
\acmYear{2022}
\setcopyright{acmcopyright}\acmConference[MSR '22]{19th International Conference
on Mining Software Repositories}{May 23--24, 2022}{Pittsburgh, PA, USA}
\acmBooktitle{19th International Conference on Mining Software Repositories (MSR
'22), May 23--24, 2022, Pittsburgh, PA, USA}
\acmPrice{15.00}
\acmDOI{10.1145/3524842.3528431}
\acmISBN{978-1-4503-9303-4/22/05}

\acmConference[MSR 2022]{MSR '22: Proceedings of the 19th International Conference on Mining Software Repositories}{May 23–24, 2022}{Pittsburgh, PA, USA}

\usepackage{pgfplots}
\usepackage{prettyref}
\usepackage{tikz}
\usepackage{xcolor,colortbl}
\usepackage{multirow}
\usepackage[ruled, linesnumbered, vlined, commentsnumbered]{algorithm2e}
\usepackage[section]{placeins}
\usepackage{booktabs}
\usepackage{threeparttable}
\usepackage{makecell}
\usepackage{graphicx}
\usepackage[compact]{titlesec}
\usepackage{Figures/pgf-pie}
\usepackage[listings,skins,breakable]{tcolorbox}
\usepackage{csquotes}
\usepackage{subcaption}
\usepackage{standalone}
\usepackage[skip=0.5\baselineskip]{caption}
\usepackage{wrapfig}
\usepackage{amsmath}
\usetikzlibrary{pgfplots.statistics}
\usepackage{tcolorbox}
\usepackage{caption}
\usepackage{url}
\usepackage{multirow, hhline}
\usepackage{enumitem}

\newtcolorbox{quotebox}{colback=blue!10,boxrule=0.4pt,colframe=black,fonttitle=\bfseries,top=2pt,bottom=2pt}

\newtcolorbox{resultbox}{colback=blue!10,boxrule=0.4pt,colframe=black,fonttitle=\bfseries,top=2pt,bottom=2pt}

\newtcolorbox{insightbox}{
  sidebyside,sidebyside align=top,lower separated=true,lefthand width=0.5em,
  arc=0pt,
  boxsep=0pt,
  left=2pt,right=2pt,top=2pt,bottom=2pt,colframe=white,
  skin=bicolor,
  colback=black!50, 
  colupper=white,
  colbacklower=white,boxrule=0pt,colframe=white,
  sidebyside gap=5pt,
}

\mathchardef\mhyphen="2D

\DeclareMathAlphabet\mathbfcal{OMS}{cmsy}{b}{n}

\newbox\aMark
\setbox\aMark\hbox{\begin{pgfpicture}\textcolor{red}{\pgfuseplotmark{o}}\end{pgfpicture}}

\newbox\bMark
\setbox\bMark\hbox{\begin{pgfpicture}\textcolor{red}{\pgfuseplotmark{star}}\end{pgfpicture}}

\newrefformat{fig}{Fig.~\ref{#1}}
\newrefformat{tab}{Table~\ref{#1}}
\newrefformat{sec}{Section~\ref{#1}}
\newrefformat{app}{Appendix~\ref{#1}}
\newrefformat{alg}{Algorithm~\ref{#1}}
\newrefformat{property}{Property~\ref{#1}}
\newrefformat{theorem}{Theorem~\ref{#1}}
\newrefformat{lemma}{Lemma~\ref{#1}}
\newrefformat{corollary}{Corollary~\ref{#1}}
\newrefformat{proposition}{Proposition~\ref{#1}}
\newrefformat{def}{Definition~\ref{#1}}
\newrefformat{eq}{equation~(\ref{#1})}

\definecolor{steel}{rgb}{0, 0.2, 0.9} 

\pgfplotsset{compat=newest}

\pgfplotsset{compat=1.11,
    /pgfplots/ybar legend/.style={
    /pgfplots/legend image code/.code={%
       \draw[##1,/tikz/.cd,yshift=-0.25em]
        (0cm,0cm) rectangle (3pt,0.8em);},
   },
}

\def\signed #1{{\leavevmode\unskip\nobreak\hfil\penalty50\hskip2em
  \hbox{}\nobreak\hfil(#1)%
  \parfillskip=0pt \finalhyphendemerits=0 \endgraf}}

\newsavebox\mybox

   \newcommand{\quart}[4]{\begin{adjustbox}{max width=.1\textwidth}\begin{picture}(100,5)
    {\color{black}\put(#1,5){\line(1,0){#2}}\color{blue!50}\put(#3,5){\circle*{7}}\color{black}\put(#3,5){\circle{7}}}\end{picture}\end{adjustbox}}

\begin{document}

\title[Does Configuration Encoding Matter in Learning Software Performance?]{Does Configuration Encoding Matter in Learning Software Performance? An Empirical Study on Encoding Schemes}

\author{Jingzhi Gong}
\affiliation{%
  \institution{Loughborough University}
  \city{Loughborough}
  \country{UK}
 }
\email{j.gong@lboro.ac.uk}

\author{Tao Chen}
\authornote{Corresponding Author}
\affiliation{
  \institution{Loughborough University}
  \city{Loughborough}
  \country{UK}
}
\email{t.t.chen@lboro.ac.uk}






\begin{abstract}

Learning and predicting the performance of a configurable software system helps to provide better quality assurance. One important engineering decision therein is how to encode the configuration into the model built. Despite the presence of different encoding schemes, there is still little understanding of which is better and under what circumstances, as the community often relies on some general beliefs that inform the decision in an ad-hoc manner. To bridge this gap, in this paper, we empirically compared the widely used encoding schemes for software performance learning, namely label, scaled label, and one-hot encoding. The study covers five systems, seven models, and three encoding schemes, leading to 105 cases of investigation. 

Our key findings reveal that: (1) conducting trial-and-error to find the best encoding scheme in a case by case manner can be rather expensive, requiring up to 400$+$ hours on some models and systems; (2) the one-hot encoding often leads to the most accurate results while the scaled label encoding is generally weak on accuracy over different models; (3) conversely, the scaled label encoding tends to result in the fastest training time across the models/systems while the one-hot encoding is the slowest; (4) for all models studied, label and scaled label encoding often lead to relatively less biased outcomes between accuracy and training time, but the paired model varies according to the system.

We discuss the actionable suggestions derived from our findings, hoping to provide a better understanding of this topic for the community. To promote open science, the data and code of this work can be publicly accessed at
\texttt{\textcolor{blue}{\url{https://github.com/ideas-labo/MSR2022-encoding-study}}}.

\end{abstract}






\maketitle

\section{Introduction}
\label{sec:introduciton}



Configurable software systems allow software engineers to tune a set of configurations options (e.g., the \texttt{cache\_size} in \textsc{MongoDB}), which can considerably influence their performance, such as latency, runtime and energy consumption, \textit{etc}.~\cite{DBLP:journals/tosem/ChenLBY18,DBLP:journals/corr/abs-1801-02175}. This is, in fact, a two-edged sword: on one hand, these configuration options offer the flexibility for software to deal with different needs, and even create the foundation to achieve runtime self-adaptation; on the other hand, their combinatorial implications to the performance are often unclear, which may result in severe complication and consequences for software maintenance. For example,~\citeauthor{DBLP:conf/sigsoft/XuJFZPT15}~\cite{DBLP:conf/sigsoft/XuJFZPT15} have discovered that software engineers find it generally difficult to adjust the configurations options in order to adapt the performance. ~\citeauthor{DBLP:conf/esem/HanY16}~\cite{DBLP:conf/esem/HanY16} have further shown that over 59\% of the performance bugs nowadays are due to inappropriate configurations. Therefore, to take full advantage of the configurability and adaptability of the software, a performance model, which takes a possible configuration as inputs to predict the likely performance, is of high importance.

Classic performance model has been relying on analytical methods, but soon they become ineffective due primarily to the soaring complexity of modern software systems. In particular, there are two key reasons which prevent the success of analytical methods: (1) analytical models often work on a limited type of configuration options, such as CPU and memory settings~\cite{DBLP:conf/wosp/DidonaQRT15,DBLP:journals/tse/ChenB17}, which cannot cope with the increasing complexity of modern systems. For example, it has been reported that configurable software systems often contain more complex and diverse types of configuration options that span across different modules, including cache, threading, and parallelism, \textit{etc}~\cite{DBLP:conf/sigsoft/XuJFZPT15}. (2) Their effectiveness is highly dependent on assumptions about the internal structure and the environment of the software being modeled. However, many modern scenarios, such as cloud-based systems, virtualized and multi-tenant software, intentionally hide such information to promote ease of use, which further reduces the reliability of the analytical methods~\cite{DBLP:conf/icse/Chen19b}. To overcome the above, machine learning based performance modelings have been gaining momentum in recent years~\cite{DBLP:journals/software/KalteneckerGSA20}, as they require limited assumption, work on arbitrary types of configurations options, and do not rely on heavy human intervention.

A critical engineering decision to make in learning performance for configurable software is how to encode the configurations. In the literature, three encoding schemes are prevalent: (1) embedding the configuration options without scaling (label encoding)~\cite{DBLP:journals/corr/abs-1801-02175,DBLP:conf/sigsoft/SiegmundGAK15,DBLP:conf/sigsoft/NairMSA17}; (2) doing so with normalization (scaled label encoding)~\cite{DBLP:conf/icse/Chen19b,DBLP:journals/tse/ChenB17,DBLP:conf/icse/HaZ19} or (3) converting them into binary ones that focus on the configuration values of those options, each of which serves as a dimension (one-hot encoding)~\cite{DBLP:conf/icse/SiegmundKKABRS12,DBLP:conf/kbse/GuoCASW13,DBLP:conf/kbse/BaoLWF19}. 


Existing work takes one of these three encoding schemes without systematic justification or even discussions, leaving us with little understanding in this regard. This is of concern, as in other domains, such as system security analysis~\cite{9020560} and medical science~\cite{DBLP:journals/bmcbi/HeP16}, it has been shown that the encoding scheme chosen can pose significant implications to the success of a machine learning model. Further, choosing one in a trial-and-error manner for each case can be impractical and time-consuming, as we will show in Section~\ref{sec: analysis and results}. It is, therefore, crucial to understand how the encoding performs differently for learning performance of configurable software.

To provide a better understanding of this topic, in this paper, we conduct an empirical study that systematically compares the three encoding schemes for learning software performance and discuss the insights learned. Our hope is to provide more justified understandings towards such an engineering decision in learning software performance under different circumstances. 



\subsection{Research Questions}

Our study covers seven widely used machine learning models for learning software performance, i.e., Decision Tree (DT)~\cite{DBLP:series/smpai/RokachM14} (used by~\cite{DBLP:conf/icse/Chen19b,DBLP:journals/tse/ChenB17,DBLP:journals/corr/abs-1801-02175,DBLP:conf/kbse/GuoCASW13}), $k$-Nearest Neighbours ($k$NN)~\cite{fix1985discriminatory} (used by ~\cite{DBLP:journals/software/KalteneckerGSA20}), Kernel Ridge Regression (KRR)~\cite{vovk2013kernel} (used by~\cite{DBLP:journals/software/KalteneckerGSA20}), Linear Regression (LR)~\cite{DBLP:journals/tamm/Goldin10} (used by~\cite{DBLP:conf/icse/Chen19b,DBLP:journals/tse/ChenB17,DBLP:conf/icse/SiegmundKKABRS12}), Neural Network (NN)~\cite{wang2003artificial} (used by~\cite{DBLP:conf/icse/HaZ19, fei2016compressor}), Random Forest (RF)~\cite{DBLP:conf/icdar/Ho95} (used by~\cite{DBLP:conf/splc/ValovGC15,DBLP:conf/oopsla/QueirozBC16}), and Support Vector Regression (SVR)~\cite{cortes1995support} (used by~\cite{DBLP:conf/icse/Chen19b,DBLP:conf/splc/ValovGC15}), together with five popular real-world software systems from prior work~\cite{DBLP:journals/corr/abs-1801-02175, DBLP:journals/corr/abs-2106-02716,ChenMMO21,DBLP:journals/corr/abs-2112-07303}, covering a wide spectrum of characteristics and domains. Naturally, the first research question (RQ) we ask is:

\begin{quotebox}
   \noindent
   \textit{\textbf{RQ1:} Is it practical to examine all encoding methods for finding the best one under every system?}
\end{quotebox}

\textbf{RQ1} seeks to confirm the significance of our study: if it takes an unreasonably long time to conduct trial-and-error in a case-by-case manner, then guidelines on choosing the best encoding scheme under different circumstances become rather important. 


What we seek to understand next is:

\begin{quotebox}
   \noindent
   \textit{\textbf{RQ2:} Which encoding scheme (paired with the model) helps to build a more accurate performance model?}
\end{quotebox}

We use Root Mean Squared Error (RMSE), which is commonly used for software performance modeling~\cite{DBLP:conf/mascots/GrohmannSELKD20,DBLP:conf/cloudcom/IorioHTA19}, as the metric for accuracy. In particular, to make a comparison under the best possible situation, we follow the standard pipeline in software performance learning~\cite{DBLP:conf/icse/Chen19b,DBLP:journals/tse/ChenB17,DBLP:journals/corr/abs-1801-02175,DBLP:conf/sigsoft/SiegmundGAK15,DBLP:conf/sigsoft/NairMSA17} that tunes the hyperparameters of each model-encoding pair using grid search and cross-validation, which is a common way for parameter tuning~\cite{DBLP:series/lncs/Hinton12}. 

While prediction accuracy is important, the time taken for training can also become an integral factor in software performance learning. Our next RQ is, therefore:


\begin{quotebox}
   \noindent
   \textit{\textbf{RQ3:} Which encoding scheme (paired with the model) helps train a performance model faster?}
\end{quotebox}

We examine the training time of each model-encoding pair, including all processes in the learning pipeline such as hyperparameter tuning and validation.


Since it is important to understand the relationship between accuracy and training time, in the final RQ, we ask:

\begin{quotebox}
   \noindent
   \textit{\textbf{RQ4:} What are the trade-offs between accuracy and training time when choosing the encoding and models?}
\end{quotebox}

With this, we seek to understand the Pareto-optimal choices that are neither the highest on accuracy nor has the fastest training time (the non-extreme points), especially those that achieve a well-balanced between accuracy and training time, i.e., the knee points.

\subsection{Contributions}

In a nutshell, we show that choosing the encoding scheme is non-trivial for learning software performance and our key findings are:

\begin{itemize}
    \item  \textbf{To RQ1:} Performing trial-and-error in a case by case manner for finding the best encoding schemes can be rather expensive under some cases, e.g., up to 400$+$ hours.
    \item  \textbf{To RQ2:} The one-hot and label encoding tends to be the best choice while the scaled label encoding performs generally the worst.
    \item  \textbf{To RQ3:} Opposed to \textbf{RQ2}, the scaled label encoding is generally the best choice while the one-hot encoding often exhibit the slowest training.
    \item  \textbf{To RQ4:} Over the models studied, the label and scaled label encoding often lead to less biased results, particularly the latter, but the paired model varies depending on the system.
\end{itemize}

Deriving from the above, we provide actionable suggestions for learning software performance under a variety of circumstances:

\begin{enumerate}
    \item When the model to be used involves RF, SVR, KRR, or NN, it is recommended to avoid trial-and-error for finding the best encoding schemes. However, this may be practical for $k$NN, DT, and LR.
    \item When the accuracy is of primary concern, 
    \begin{itemize}
        \item using neural network paired with one-hot encoding if all models studied are available to choose.
        \item using one-hot encoding for deep learning (NN), lazy models ($k$NN), and kernel models (KRR and SVM).
        \item using label encoding for linear (LR) and tree models (DT and RF).
    \end{itemize}
    \item When the training time is more important,
     \begin{itemize}
        \item using linear regression paired with scaled label encoding if all models studied are available to choose.
        \item using scaled label encoding for deep learning (NN), linear (LR), and kernel models (KRR and SVR).
        \item using label encoding for lazy ($k$NN) and tree models (DT and RF).
    \end{itemize}
    \item When a trade-off between accuracy and training time is unclear while an unbiased outcome is preferred, 
     \begin{itemize}
        \item using scaled label encoding for achieving a relatively well-balanced result if considering all models studied, but the paired model requires some efforts to determine.
        \item if the model is fixed, only the kernel models (KRR and SVR) and lazy model ($k$NN) have a more balanced outcome achieved by label encoding and scaled label encoding, respectively.
    \end{itemize}
\end{enumerate}

The remaining of this paper is organized as follows: Section~\ref{sec:pre} introduces the background information. Section~\ref{sec:methodology} elaborates the details of our empirical strategy. Section~\ref{sec: analysis and results} discusses the results and answers the aforementioned research questions. The insights learned and actionable suggestions are specified in Section~\ref{sec: discussions}. Section~\ref{sec:new-discussions} discusses the implications of our study. Section~\ref{sec:tov},~\ref{sec:related}, and~\ref{sec: conclusions} present the threats to validity, related work, and conclusion, respectively.

\section{Preliminaries}
\label{sec:pre}


In this section, we elaborate on the necessary background information and the motivation of our study.

\subsection{Learning Software Performance}

A configurable software system comes with several configuration options, such as the \texttt{interval} for \textsc{MongoDB}. Each of these options can be configured using a set of predefined values, and therefore they are often treated as discrete values, including binary, categorical or numeric options, e.g., we may set $(1,2,3,4)$ on the \texttt{interval}.

\begin{table}[t!]
\centering
\footnotesize
\caption{An example of configurations and performance for \textbf{MongoDB}. $x_i$ is the $i$th configuration option and $\varmathbb{P}$ is the performance value (runtime).}
\begin{tabular}{lllllll||c}
\toprule
$x_{1}$ & $x_{2}$ & $x_{3}$ & $\cdots$ & $x_{n-2}$ & $x_{n-1}$ & $x_{n}$ & $\varmathbb{P}$ \\ \midrule
0 & 0 & 0 & $\cdots$ & 0 & 3 & 10 & 1200 seconds \\ 
0 & 1 & 0 & $\cdots$ & 0 & 2 & 11 & 2100 seconds \\ 
$\cdots$        & $\cdots$  & $\cdots$       & $\cdots$       & $\cdots$       & $\cdots$ & $\cdots$ & $\cdots$ \\ 
0 & 0 & 1 & $\cdots$ & 0 & 9 & 23 & 1260 seconds \\ 
0 & 0 & 1 & $\cdots$ & 1 & 8 & 65 & 1140 seconds \\ \bottomrule
\end{tabular}
\label{tb:data_example}
\end{table}

Without loss of generality, as shown in Table~\ref{tb:data_example}, learning performance for a configurable software often aims to build a regression model that predicts a performance attribute $\varmathbb{P}$~\cite{DBLP:journals/tse/ChenB17,DBLP:conf/icse/ChenB13,DBLP:conf/ucc/ChenBY14}, e.g., runtime, written as: 
\begin{equation}
    \varmathbb{P} = f(\mathbf{\overline{x}})\text{, } f : \mathbf{\overline{x}} \rightarrow \mathcal{R}
\end{equation}
whereby $f$ is the actual function learned by a machine learning model; $\mathbf{\overline{x}}$ is the vector that represents a configuration. Given that configurable software runs under an environment, the aim is to train a model that minimizes the generalization error on new configurations which have not been seen in training.

\subsection{Encoding Schemes}

In machine learning, the steps involved in the automated model building forms a \textbf{\textit{learning pipeline}}~\cite{mohr2021predicting}. For learning software performance, the standard learning pipeline setting consists of preprocessing, hyperparameter tuning, model training (using all configuration options), and model evaluation~\cite{DBLP:conf/icse/Chen19b,DBLP:journals/tse/ChenB17,DBLP:journals/corr/abs-1801-02175,DBLP:conf/sigsoft/SiegmundGAK15,DBLP:conf/sigsoft/NairMSA17} (see Section~\ref{sec:methodology} for details).

In all learning pipeline phases, one critical engineering decision, which this paper focuses on, is how the $\mathbf{\overline{x}}$ can be encoded. In general, existing work takes one of the following three encoding schemes:

   \textbf{Label Encoding:} This is a widely used scheme~\cite{DBLP:journals/corr/abs-1801-02175,DBLP:conf/sigsoft/SiegmundGAK15,DBLP:conf/sigsoft/NairMSA17}, where each of the configuration options occupies one dimension in $\mathbf{\overline{x}}$. Taking \textsc{MongoDB} as an example, its configuration can be represented as $\mathbf{\overline{x}}=($\texttt{cache\_size}$,$ \texttt{interval}$,$ \texttt{ssl}$,$ \texttt{data\_strategy}$)$ where \texttt{cache\_size}$=(1,10,10000)$, \texttt{interval}$=(1,2,3,4)$, \texttt{ssl}$=(0,1)$ and \texttt{data\_strategy}$=(str\_l1,$ $str\_l2,str\_l3)$. A configuration that is used as a training sample could be $(10000,2,1,1)$, where the $\texttt{data\_}$$\texttt{strategy}$ can be converted into numeric values of $(0,1,2)$.
    
     \textbf{Scaled Label Encoding:} This is a variant of the label encoding used by a state-of-the-art approach~\cite{DBLP:conf/icse/Chen19b,DBLP:journals/tse/ChenB17,DBLP:conf/icse/HaZ19}, where each configuration also takes one dimension in $\mathbf{\overline{x}}$. The only difference is that all configurations are normalized to the range between 0 and 1. The same example configuration above for label encoding would be scaled to $(1, \frac{1}{3}, 1, 0.5)$.
    
  \textbf{One-hot Encoding:} Another commonly followed scheme~\cite{DBLP:conf/icse/SiegmundKKABRS12,DBLP:conf/kbse/GuoCASW13,DBLP:conf/kbse/BaoLWF19} such that each dimension in $\mathbf{\overline{x}}$ refers to the binary form of a particular value for a configuration option. Using the above example of \textsc{MongoDB}, the representation becomes $\mathbf{\overline{x}}=($\texttt{cache\_size\_v1}$,$ \texttt{cache\_size\_v2}$,...)$. Each dimension, e.g., \texttt{cache\_size\_v1}, would have a value of 1 if it is the one that the corresponding configuration chooses, otherwise it is 0. As such, the same configuration $(10000,2,1,1)$ in the label encoding would be represented as $(0,0,1,0,1,0,0,0,1,0,1,0)$ in the one-hot encoding.


Clearly, for binary options, the three encoding methods would be identical, hence in this work, we focus on the systems that also come with complex numeric and categorical configuration options.

\subsection{Why Study Them?}

Despite the prevalence of the three encoding schemes, existing work often use one of them without justifying their choice for learning software performance~\cite{DBLP:conf/icse/Chen19b,DBLP:journals/tse/ChenB17,DBLP:journals/corr/abs-1801-02175,DBLP:conf/sigsoft/SiegmundGAK15,DBLP:conf/sigsoft/NairMSA17,DBLP:conf/icse/SiegmundKKABRS12,DBLP:conf/kbse/GuoCASW13,DBLP:conf/kbse/BaoLWF19}, particularly relating to the accuracy and training time required for the model. Some studies have mentioned the rationals, but a common agreement on which one to use has not yet been drawn. For example, \citeauthor{DBLP:conf/kbse/BaoLWF19}~\cite{DBLP:conf/kbse/BaoLWF19} state that for categorical configuration options, e.g., \texttt{cache\_mode} with three values (\texttt{memory}, \texttt{disk}, \texttt{mixed}), the label encoding unnecessarily assume a natural ordering between the values, as they are represented as $1$, $2$, and $3$. Here, one-hot encoding should be chosen. However, \citeauthor{DBLP:journals/jmlr/AlayaBGG19}~\cite{DBLP:journals/jmlr/AlayaBGG19} argue that the one-hot encoding can easily suffer from the multicollinearity issue on categorical configuration options, i.e., it is difficult to handle options interaction. For numeric configuration options, label encoding may fit well, as it naturally comes with order, e.g., the \texttt{cache\_size} in \textsc{MongoDB}, which has a set of values $(1,10,10000)$. However, the values, such as the above, can be of largely different scales and thus degrade numeric stability. Indeed, using one-hot encoding could be robust to this issue, but it loses the ordinal property of the numeric configuration option~\cite{DBLP:conf/sigsoft/SiegmundGAK15}. Similarly, scaled label encoding could reduce the instability and improve the prediction performance~\cite{pan2016impact}, but it also weakens the interactions between the scaled options and the binary options (as they stay the same). Therefore, there is still no common agreement (or insights) on which encoding scheme to use under what circumstances for learning performance models.


Unlike other domains, software configuration is often highly sparse, leading to unusual data distributions. Specifically, a few configuration options could have large influence on the software performance, while the others are trivial, which makes the decision of encoding scheme difficult. Moreover, it is often the case that we may not fully understand the nature of every configuration option, as the software may be off-the-shelf or close-sourced; hence, one may not be able to choose the right encoding based on purely theoretical understandings. As such, a high-level guideline on choosing the encoding scheme for performance modeling, which gives overall suggestions for the practitioners, is in high demand. 

The above thus motivates this empirical study, aiming to analyzing the effectiveness of encoding schemes across various subject systems and machine learning models, summarizing the common behaviors of the encoding methods, and providing actionable advice based on the learning models applied as well as the requirements, e.g., accuracy and training time.


\section{Methodology}
\label{sec:methodology}

In this section, we will discuss the methodology and experimental setup of the empirical strategy for our study.


\subsection{System and Data Selection}
We set the following criteria to select sampled data of configurable systems and their environments when comparing the three encoding schemes:

\begin{enumerate}

\item To promote the reproducibility,
the systems should be open-sourced and the data should be hosted in public repositories, e.g., GitHub.

\item The system and its environment should have been widely used and tested in existing work.

\item To ensure a case where the encoding schemes can create sufficiently different representations, the system should have at least 10\% configuration options that are not categorical/binary.

\item To promote the robustness of our experiments, the subject systems should have different proportions of configuration options that are numerical.

\item To guarantee the scale of the study, we consider systems with more than 5,000 configuration samples.

\end{enumerate}

We shortlisted systems and their data from recent studies on software configuration tuning and modeling~\cite{DBLP:journals/corr/abs-1801-02175, DBLP:journals/corr/abs-2106-02716}, from which we identified five systems and their environment according to the above criteria, as shown in Table~\ref{tb:subj}. The five systems contain different percentages of categorical/binary and numeric configuration options while covering five distinct domains. 

Note that since the measurement and sampling process for configurable software is usually rather expensive, e.g., \citeauthor{DBLP:conf/icml/ZuluagaSKP13}~\cite{DBLP:conf/icml/ZuluagaSKP13} report that the synthesis of only one software configuration design can take hours or even days, in practice it is not necessarily always possible to gather an extremely large number of data samples. Further, using the full samples for some systems with a large configuration space can easily lead to unrealistic training time for certain models, e.g., with Neural Network, it took several days to complete only one run under our learning pipeline on the full datasets of \textsc{Trimesh}. Therefore in this work, for each system, we randomly sample 5,000 configurations from the dataset as our experiment data, which tends to be reasonable and is also a commonly used setting in previous work~\cite{DBLP:conf/kbse/DornAS20,DBLP:conf/im/JohnssonMS19, DBLP:conf/icdm/ShaoWL19, DBLP:conf/icse/Gerostathopoulos18}.


\begin{table}
  \caption{Datasets of configurable software systems used. $\lvert \mathbfcal{O} \rvert$ (C/N) denotes the number of categorical (including binary) / numerical options.}
  \footnotesize
  \setlength{\tabcolsep}{1.4mm} 
  \label{tb:subj}
  \begin{tabular}{ccccc}
    \toprule
    \textbf{Dataset}&$\lvert \mathbfcal{O} \rvert$ (C/N)& \textbf{Performance} & \textbf{Description}&\textbf{Used by}\\
    \midrule
    \textsc{MongoDB} & 14/2 & runtime (ms) & NoSQL database & \cite{DBLP:journals/corr/abs-2106-02716}\\
    
    \textsc{Lrzip} & 9/3 & runtime (ms) &  compression tool & \cite{DBLP:journals/corr/abs-2106-02716}\\
    
    \textsc{Trimesh} & 9/4 & runtime (ms) & triangle meshes library& \cite{DBLP:journals/corr/abs-1801-02175,ChenMMO21}\\

    \textsc{ExaStencils} & 4/6 & latency (ms) & stencil code generator& \cite{DBLP:journals/corr/abs-2106-02716}\\
    
    \textsc{x264} & 4/13 & energy (mW) & a video encoder & \cite{DBLP:journals/corr/abs-1801-02175,ChenMMO21}\\
  \bottomrule
\end{tabular}

\end{table}

\subsection{Machine Learning Models}


In this work, we choose the most common models that are of different types as used in prior studies:

\begin{itemize}

\item\textbf{Linear Model:} This type of model build the correlation between configurations options and performance under certain linear assumptions. 

\begin{itemize}
    \item\textbf{Linear Regression (LR):} A multi-variable linear regression model that linearly correlate the configurations and their options to make prediction. It has been used by~\cite{DBLP:conf/icse/Chen19b,DBLP:journals/tse/ChenB17,DBLP:conf/icse/SiegmundKKABRS12}. There are three hyperparameters to tune, e.g., \texttt{n\_jobs}.
\end{itemize}

\item\textbf{Deep Learning Model:} A model that is based on multiple layers of perceptrons to learn and predict the concepts. 

\begin{itemize}

\item\textbf{Neural Network (NN):} A network structure with layers of neurons and connections representing the flow of data. The weights incorporate the influences of each input unit and interactions between them. The NN models have been shown to be successful for modeling software performance, e.g.,~\cite{DBLP:conf/icse/HaZ19, fei2016compressor}. In this work, we utilize the same network setting and hyperparameter tuning method from the work by \citeauthor{DBLP:conf/icse/HaZ19}~\cite{DBLP:conf/icse/HaZ19}. 

\end{itemize}

\item\textbf{Tree Model:} This model constructs a tree-like structure with a clear decision boundary on the branches.

\begin{itemize}
    \item\textbf{Decision Tree (DT):} A regression tree model that recursively partitions the configurations space to predict the output, which is used by~\cite{DBLP:conf/icse/Chen19b,DBLP:journals/tse/ChenB17,DBLP:journals/corr/abs-1801-02175,DBLP:conf/kbse/GuoCASW13}. We tune a DT using three hyperparameters, e.g., \texttt{min\_samples\_split}.

\item\textbf{Random Forest (RF):} An ensemble of decision trees, each of which learns a subset of samples or configurations space. It is a widely used model for performance learning~\cite{DBLP:conf/splc/ValovGC15,DBLP:conf/oopsla/QueirozBC16}. There are three hyperparameters to tune, e.g., \texttt{n\_estimators}

\end{itemize}

\item\textbf{Lazy Model:} This model delays the learning until the point of prediction.

\begin{itemize}

\item\textbf{$k$-Nearest Neighbours ($k$NN):} A model that considers only already measured neighboring configurations to produce a prediction, which has been commonly used~\cite{DBLP:journals/software/KalteneckerGSA20}. It has four hyperparameters to be tuned, such as \texttt{n\_neighbors}.

\end{itemize}

\item\textbf{Kernel Model:} This model performs learning and prediction by means of a kernel function.

\begin{itemize}
\item\textbf{Kernel Ridge Regression (KRR):} A model of kernel transformation that is combined with ridge
regression, which is the $L_2$-norm regularization. It has been used by~\cite{DBLP:journals/software/KalteneckerGSA20}.  There are three hyperparameters to tune, such as \texttt{alpha}.

\item\textbf{Support Vector Regression (SVR):} A model that transforms the configurations space into a higher- dimensional space via the kernel function, as used by~\cite{DBLP:conf/icse/Chen19b,DBLP:conf/splc/ValovGC15}. It contains hyperparameters, e.g., \texttt{kernel\_func}.
\end{itemize}

\end{itemize}

The reasons for the choices are two-fold: (1) they are prominently used in previous work; and (2) they exist standard implementation under the same and widely-used machine learning library, i.e., \texttt{Sklearn}~\cite{DBLP:journals/jmlr/PedregosaVGMTGBPWDVPCBPD11} and \texttt{Tensorflow}, which reduces the possibility of bias. Note that we did not aim to be exhaustive, but focusing on those that are the most prevalent ones such that the potential impact of this study can be maximized.

\subsection{Metrics}

Different metrics exist for measuring the accuracy of a prediction model. In this study, we use RMSE because of two reasons: (1) it is a widely used metric for performance modeling of configurable systems in prior work~\cite{DBLP:conf/mascots/GrohmannSELKD20,DBLP:conf/cloudcom/IorioHTA19}; and (2) it has been reported that RMSE can reveal the performance difference better, compared with its popular counterparts such as Mean Relative Error~\cite{chai2014root}. RMSE is calculated as:
\begin{equation}
    RMSE = \sqrt{ \frac{1}{N}\sum_{i=1}^{N} (x_{i} - \hat{x}_{i})^2}
\end{equation}
whereby $x_{i}$ and $\hat{x}_{i}$ are the actual and predicted performance value, respectively; $N$ denotes that total number of testing data samples.

As for the training time, we report the time taken for completing the training process, including hyperparameter tuning and prepossessing as necessary.

\subsection{Learning Pipeline Setting}

\begin{figure}[t!]
\centering
\includegraphics[width=\columnwidth]{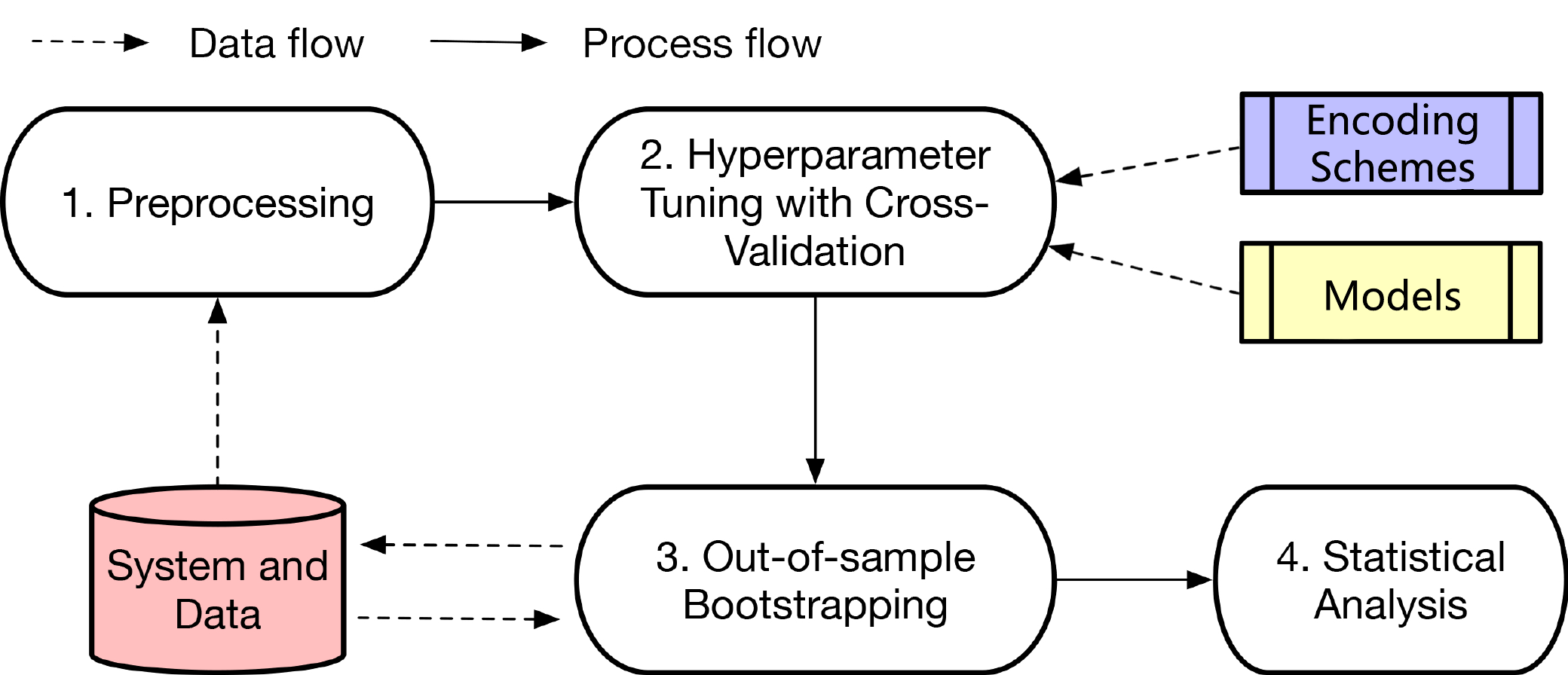}
\caption{The learning pipeline in this study.}
\label{fig:wf}
\end{figure}

As shown in Figure~\ref{fig:wf}, the standard learning pipeline setting in our empirical study has several key steps as specified below:

\begin{enumerate}[leftmargin=0.2in]
    \item\textbf{Preprocessing:} For label and one-hot encoding, we utilize the standard encoding functions from the \texttt{Sklearn} library. For the scaled label encoding, we normalize the configurations using the \textit{max-min scaling}, such that an option value $v$ is standardized as $v= {{v - v_{min}} \over {v_{max}- v_{min}}}$, where $v_{max}$ and $v_{min}$ denote the maximum and minimum bound, respectively. In this way, the values of each configuration option can be normalized within the range between $0$ and $1$. We follow the state-of-the-art learning pipeline such that all configuration options and their values are considered in the model ~\cite{DBLP:journals/tse/ChenB17,DBLP:journals/corr/abs-1801-02175,DBLP:conf/sigsoft/SiegmundGAK15,DBLP:conf/sigsoft/NairMSA17,DBLP:conf/kbse/BaoLWF19}.

   \item\textbf{Hyperparameter Tuning:} It is not uncommon that a model comes with at least one hyperparameter~\cite{DBLP:conf/icse/LiX0WT20}. Therefore, the common practice of the pipeline for learning software performance is to tune them under all encoding schemes~\cite{DBLP:journals/tse/ChenB17,DBLP:journals/corr/abs-1801-02175,DBLP:conf/sigsoft/SiegmundGAK15,DBLP:conf/sigsoft/NairMSA17,DBLP:conf/kbse/BaoLWF19}. In this study, we use the \texttt{GridSearchCV} function from \texttt{Sklearn}, which is an exhaustive grid search that evaluates the model quality via 10-fold cross-validation on the training dataset. The one that leads to the best result is used. Note that the default values are always used as a starting point.

    \item\textbf{Bootstrapping:} To achieve a reliable conclusion, we conducted out-of-sample bootstrap (without replacement). In particular, we randomly sampled 90\% of the data as the training dataset, those samples that were not included in the training were used as the testing samples. The process was repeated 50 times, i.e., there are 50 runs of RMSE (on the testing dataset) and training time to be reported. For each run, all encoding schemes are examined, thereby we ensure that they are evaluated under the same randomly sampled training and testing dataset.

    \item\textbf{Statistical Analysis:} To ensure statistical significance in multiple comparisons, we apply Scott-Knott test~\cite{DBLP:journals/tse/MittasA13} on all comparisons of over 50 runs and produce a score. In a nutshell, Scott-Knott sorts the list of treatments (the learning model-encoding pairs) by their median RMSE/training time. Next, it splits the list into two sub-lists with the largest expected difference~\cite{xia2018hyperparameter}. Suppose that we compare \texttt{NN\_onehot}, \texttt{RF\_onehot}, and \texttt{NN\_label}, a possible split could be: $\{$\texttt{NN\_onehot}, \texttt{RF\_onehot}$\}$, $\{$\texttt{NN\_label}$\}$, with the score of 2 and 1, respectively. This means that, in the statistical sense, \texttt{NN\_onehot}~and \texttt{RF\_onehot} perform similarly, but they are significantly better than \texttt{NN\_label}. Formally, Scott-Knott test aims to find the best split by maximizing the difference $\Delta$ in the expected mean before and after each split:
\begin{equation}
    \Delta = \frac{|l_1|}{|l|}(\overline{l_1} - \overline{l})^2 + \frac{|l_2|}{|l|}(\overline{l_2} - \overline{l})^2
\end{equation}
whereby $|l_1|$ and $|l_2|$ are the sizes of two sub-lists ($l_1$ and $l_2$) from list $l$ with a size $|l|$. $\overline{l_1}$, $\overline{l_2}$, and $\overline{l}$ denote their mean RMSE/training time values.

During the splitting, we apply a statistical hypothesis test $H$ to check if $l_1$ and $l_2$ are significantly different. This is done by using bootstrapping and $\hat{A}_{12}$~\cite{Vargha2000ACA}. If that is the case, Scott-Knott recurses on the splits. In other words, we divide the treatments into different sub-lists if both bootstrap sampling and effect size test suggest that a split is statistically significant (with a confidence level of 99\%) and not a small effect ($\hat{A}_{12} \geq 0.6$). The sub-lists are then scored based on their mean RMSE/training time. The higher the score, the better the treatment.
    
\end{enumerate}



Since there are five systems and environments, together with seven models and three encoding schemes, our empirical study consists of 105 cases of investigation. All the experiments were performed on a Windows 10 server with an Intel Core i5-9400 CPU 2.90GHz and 8GB RAM. 


\section{Analysis and Results}
\label{sec: analysis and results}
In this section, we discuss the results of the empirical study with respect to the RQs. All data and code can be accessed at the github repository: \texttt{\textcolor{blue}{\url{https://github.com/ideas-labo/MSR2022-encoding-study}}}.

\begin{table*}[t!]
      \caption{Scott-Knott test, Med (median), and Interquartile Range (IQR) on the RMSE of all models and systems. ``onehot'', ``label'' and ``scaled'' stand for one-hot, label and scaled label encoding, respectively. A higher score means the RMSE is lower and, therefore, better. For (a) to (e), the pairs are sorted by score, median, and then IQR. For (f), \setlength{\fboxsep}{1.5pt}\colorbox{red!10}{red} highlights the best encoding scheme for a model over all systems.}
    \label{tb:rq2-plot}
  \begin{center}
    \begin{adjustbox}{max width = \textwidth}
\footnotesize
    \begin{tabular}{c@{}c@{}c}
        \begin{tabular}{lcccl}
           \cellcolor[gray]{1}\textbf{Pair} & \cellcolor[gray]{1}\textbf{Score} &     \cellcolor[gray]{1}\textbf{Med} & \cellcolor[gray]{1}\textbf{IQR} & \cellcolor[gray]{1}\\
            \hline
\texttt{NN\_label} & 10 & 2943.38 & 0.30 & \quart{6.57}{0.30}{6.73}{100.00} \\
\texttt{NN\_scaled} & 10 & 2944.45 & 0.39 & \quart{6.52}{0.39}{6.73}{100.00} \\
\texttt{NN\_onehot} & 10 & 2951.74 & 0.31 & \quart{6.64}{0.31}{6.75}{100.00} \\
\texttt{KRR\_onehot} & 9 & 3014.68 & 0.19 & \quart{6.78}{0.19}{6.89}{100.00} \\
\texttt{RF\_label} & 8 & 3019.52 & 0.20 & \quart{6.79}{0.20}{6.91}{100.00} \\
\texttt{LR\_label} & 8 & 3019.55 & 0.19 & \quart{6.80}{0.19}{6.91}{100.00} \\
\texttt{RF\_scaled} & 8 & 3019.63 & 0.20 & \quart{6.79}{0.20}{6.91}{100.00} \\
\texttt{LR\_onehot} & 8 & 3020.36 & 0.21 & \quart{6.79}{0.21}{6.91}{100.00} \\
\texttt{KRR\_label} & 8 & 3024.03 & 0.19 & \quart{6.81}{0.19}{6.92}{100.00} \\
\texttt{LR\_scaled} & 8 & 3026.49 & 0.19 & \quart{6.83}{0.19}{6.92}{100.00} \\
\texttt{KRR\_scaled} & 8 & 3028.21 & 0.19 & \quart{6.83}{0.19}{6.93}{100.00} \\
\texttt{RF\_onehot} & 7 & 3060.11 & 0.20 & \quart{6.86}{0.20}{7.00}{100.00} \\
\texttt{DT\_scaled} & 6 & 3093.12 & 0.21 & \quart{7.00}{0.21}{7.07}{100.00} \\
\texttt{DT\_onehot} & 6 & 3095.11 & 0.25 & \quart{7.00}{0.25}{7.08}{100.00} \\
\texttt{DT\_label} & 6 & 3097.61 & 0.22 & \quart{6.98}{0.22}{7.08}{100.00} \\
\texttt{$k$NN\_scaled} & 5 & 8968.09 & 1.54 & \quart{19.90}{1.54}{20.51}{100.00} \\
\texttt{$k$NN\_onehot} & 4 & 15772.52 & 0.85 & \quart{35.78}{0.85}{36.07}{100.00} \\
\texttt{$k$NN\_label} & 3 & 22114.97 & 0.99 & \quart{49.96}{0.99}{50.57}{100.00} \\
\texttt{SVR\_onehot} & 2 & 41917.30 & 5.73 & \quart{92.16}{5.73}{95.86}{100.00} \\
\texttt{SVR\_scaled} & 1 & 42228.88 & 5.85 & \quart{92.76}{5.85}{96.57}{100.00} \\
\texttt{SVR\_label} & 1 & 42229.54 & 5.86 & \quart{92.77}{5.86}{96.57}{100.00} \\
        \end{tabular} & 
        \begin{tabular}{|lcccl|}
           \cellcolor[gray]{1}\textbf{Pair} & \cellcolor[gray]{1}\textbf{Score} &     \cellcolor[gray]{1}\textbf{Med} & \cellcolor[gray]{1}\textbf{IQR} & \cellcolor[gray]{1}\\
            \hline
\texttt{NN\_onehot} & 12 & 2497.07 & 0.10 & \quart{0.61}{0.10}{0.67}{100.00} \\
\texttt{NN\_scaled} & 11 & 2887.66 & 0.20 & \quart{0.72}{0.20}{0.78}{100.00} \\
\texttt{NN\_label} & 10 & 3019.11 & 0.34 & \quart{0.71}{0.34}{0.81}{100.00} \\
\texttt{DT\_label} & 9 & 15621.69 & 5.92 & \quart{2.61}{5.92}{4.19}{100.00} \\
\texttt{DT\_scaled} & 9 & 15621.77 & 5.92 & \quart{2.61}{5.92}{4.19}{100.00} \\
\texttt{RF\_scaled} & 8 & 25417.20 & 3.37 & \quart{5.50}{3.37}{6.82}{100.00} \\
\texttt{RF\_label} & 8 & 25457.00 & 3.18 & \quart{5.70}{3.18}{6.83}{100.00} \\
\texttt{RF\_onehot} & 7 & 57173.71 & 5.31 & \quart{13.27}{5.31}{15.35}{100.00} \\
\texttt{DT\_onehot} & 7 & 59610.67 & 9.35 & \quart{10.61}{9.35}{16.00}{100.00} \\
\texttt{$k$NN\_label} & 6 & 121668.37 & 3.96 & \quart{30.25}{3.96}{32.66}{100.00} \\
\texttt{$k$NN\_scaled} & 5 & 151716.91 & 3.54 & \quart{38.37}{3.54}{40.73}{100.00} \\
\texttt{$k$NN\_onehot} & 4 & 168726.75 & 2.81 & \quart{43.77}{2.81}{45.30}{100.00} \\
\texttt{KRR\_onehot} & 3 & 192199.55 & 3.73 & \quart{49.63}{3.73}{51.60}{100.00} \\
\texttt{LR\_onehot} & 2 & 193658.25 & 3.88 & \quart{49.99}{3.88}{51.99}{100.00} \\
\texttt{LR\_scaled} & 2 & 194094.27 & 3.76 & \quart{50.36}{3.76}{52.11}{100.00} \\
\texttt{KRR\_scaled} & 2 & 194588.83 & 3.84 & \quart{50.46}{3.84}{52.24}{100.00} \\
\texttt{LR\_label} & 2 & 194648.94 & 3.79 & \quart{50.50}{3.79}{52.26}{100.00} \\
\texttt{KRR\_label} & 2 & 195128.40 & 3.88 & \quart{50.53}{3.88}{52.39}{100.00} \\
\texttt{SVR\_label} & 1 & 321782.26 & 6.84 & \quart{83.61}{6.84}{86.39}{100.00} \\
\texttt{SVR\_onehot} & 1 & 321878.92 & 6.86 & \quart{83.64}{6.86}{86.42}{100.00} \\
\texttt{SVR\_scaled} & 1 & 321923.45 & 6.86 & \quart{83.65}{6.86}{86.43}{100.00} \\
        \end{tabular} &
                \begin{tabular}{lcccl}
             \cellcolor[gray]{1}\textbf{Pair} & \cellcolor[gray]{1}\textbf{Score} &     \cellcolor[gray]{1}\textbf{Med} & \cellcolor[gray]{1}\textbf{IQR} & \cellcolor[gray]{1}
            \\
            \hline
            \texttt{NN\_scaled} & 12 & 398.39 & 8.37 & \quart{17.59}{8.37}{20.93}{100.00} \\
\texttt{NN\_label} & 12 & 398.39 & 8.37 & \quart{17.59}{8.37}{20.93}{100.00} \\
\texttt{NN\_onehot} & 11 & 645.28 & 9.91 & \quart{27.93}{9.91}{33.90}{100.00} \\
\texttt{RF\_label} & 10 & 893.39 & 9.60 & \quart{42.73}{9.60}{46.93}{100.00} \\
\texttt{RF\_scaled} & 10 & 896.18 & 9.25 & \quart{42.81}{9.25}{47.08}{100.00} \\
\texttt{RF\_onehot} & 9 & 947.02 & 11.43 & \quart{46.49}{11.43}{49.75}{100.00} \\
\texttt{DT\_onehot} & 8 & 1104.22 & 12.32 & \quart{51.41}{12.32}{58.01}{100.00} \\
\texttt{DT\_label} & 7 & 1140.59 & 17.93 & \quart{53.16}{17.93}{59.92}{100.00} \\
\texttt{DT\_scaled} & 7 & 1140.63 & 17.93 & \quart{53.16}{17.93}{59.92}{100.00} \\
\texttt{$k$NN\_onehot} & 6 & 1302.37 & 7.61 & \quart{63.67}{7.61}{68.42}{100.00} \\
\texttt{$k$NN\_scaled} & 5 & 1359.37 & 7.14 & \quart{68.06}{7.14}{71.41}{100.00} \\
\texttt{KRR\_onehot} & 4 & 1378.64 & 8.30 & \quart{68.47}{8.30}{72.42}{100.00} \\
\texttt{KRR\_scaled} & 3 & 1410.46 & 9.14 & \quart{70.32}{9.14}{74.10}{100.00} \\
\texttt{KRR\_label} & 3 & 1411.00 & 9.15 & \quart{70.34}{9.15}{74.12}{100.00} \\
\texttt{LR\_onehot} & 3 & 1411.28 & 10.07 & \quart{69.94}{10.07}{74.14}{100.00} \\
\texttt{LR\_scaled} & 3 & 1411.56 & 9.14 & \quart{70.39}{9.14}{74.15}{100.00} \\
\texttt{LR\_label} & 3 & 1412.19 & 9.08 & \quart{70.45}{9.08}{74.19}{100.00} \\
\texttt{$k$NN\_label} & 2 & 1502.74 & 10.07 & \quart{73.73}{10.07}{78.94}{100.00} \\
\texttt{SVR\_onehot} & 1 & 1521.08 & 11.72 & \quart{75.65}{11.72}{79.91}{100.00} \\
\texttt{SVR\_scaled} & 1 & 1525.33 & 11.75 & \quart{75.84}{11.75}{80.13}{100.00} \\
\texttt{SVR\_label} & 1 & 1525.57 & 11.74 & \quart{75.85}{11.74}{80.14}{100.00} \\
        \end{tabular} \\
        \\
         \textbf{\small (a). \textsc{MongoDB}} & \small \textbf{(b). \textsc{Lrzip}} & \textbf{\small (c). \textsc{Trimesh}}
        \\
        \\
        \begin{tabular}{lcccl}
            \cellcolor[gray]{1}\textbf{Pair} & \cellcolor[gray]{1}\textbf{Score} &     \cellcolor[gray]{1}\textbf{Med} & \cellcolor[gray]{1}\textbf{IQR} & \cellcolor[gray]{1}\\
            \hline
\texttt{NN\_onehot} & 14 & 67.62 & 0.88 & \quart{3.18}{0.88}{3.38}{100.00} \\
\texttt{NN\_label} & 13 & 86.18 & 0.61 & \quart{4.05}{0.61}{4.31}{100.00} \\
\texttt{NN\_scaled} & 13 & 92.37 & 1.60 & \quart{3.97}{1.60}{4.62}{100.00} \\
\texttt{RF\_scaled} & 12 & 158.61 & 4.00 & \quart{6.63}{4.00}{7.93}{100.00} \\
\texttt{RF\_label} & 12 & 159.83 & 4.00 & \quart{6.73}{4.00}{8.00}{100.00} \\
\texttt{RF\_onehot} & 12 & 164.02 & 3.99 & \quart{6.76}{3.99}{8.21}{100.00} \\
\texttt{DT\_scaled} & 11 & 179.07 & 3.46 & \quart{8.04}{3.46}{8.96}{100.00} \\
\texttt{DT\_label} & 11 & 181.96 & 3.40 & \quart{8.20}{3.40}{9.10}{100.00} \\
\texttt{DT\_onehot} & 11 & 182.42 & 3.42 & \quart{8.23}{3.42}{9.13}{100.00} \\
\texttt{$k$NN\_scaled} & 10 & 363.34 & 2.40 & \quart{17.65}{2.40}{18.18}{100.00} \\
\texttt{KRR\_onehot} & 10 & 369.58 & 0.95 & \quart{18.17}{0.95}{18.49}{100.00} \\

\texttt{$k$NN\_onehot} & 9 & 438.83 & 1.24 & \quart{21.43}{1.24}{21.95}{100.00} \\
\texttt{$k$NN\_label} & 8 & 1008.23 & 1.43 & \quart{49.62}{1.43}{50.44}{100.00} \\
\texttt{KRR\_label} & 7 & 1113.89 & 1.61 & \quart{54.82}{1.61}{55.72}{100.00} \\
\texttt{SVR\_onehot} & 6 & 1210.97 & 1.25 & \quart{59.93}{1.25}{60.58}{100.00} \\
\texttt{KRR\_scaled} & 5 & 1306.24 & 1.56 & \quart{64.59}{1.56}{65.34}{100.00} \\
\texttt{SVR\_label} & 4 & 1495.85 & 1.82 & \quart{74.05}{1.82}{74.83}{100.00} \\
\texttt{SVR\_scaled} & 3 & 1507.12 & 1.92 & \quart{74.76}{1.92}{75.39}{100.00} \\
\texttt{LR\_scaled} & 2 & 4562.24 & 1.97 & \quart{90.10}{1.97}{92.82}{100.00} \\
\texttt{LR\_label} & 2 & 4483.84 & 2.12 & \quart{90.72}{2.12}{92.45}{100.00} \\
\texttt{LR\_onehot} & 1 & 7814.81 & 99 & \quart{90.70}{9}{99.11}{100.00} \\
        \end{tabular} & 
        \begin{tabular}{|lcccl}
           \cellcolor[gray]{1}\textbf{Pair} & \cellcolor[gray]{1}\textbf{Score} &     \cellcolor[gray]{1}\textbf{Med} & \cellcolor[gray]{1}\textbf{IQR} & \cellcolor[gray]{1}\\
            \hline
\texttt{NN\_onehot} & 19 & 446.49 & 7.94 & \quart{19.60}{7.94}{22.34}{100.00} \\
\texttt{RF\_label} & 19 & 581.18 & 6.07 & \quart{27.64}{6.07}{29.07}{100.00} \\
\texttt{NN\_label} & 18 & 481.45 & 6.97 & \quart{20.77}{6.97}{24.08}{100.00} \\
\texttt{RF\_onehot} & 17 & 665.35 & 6.60 & \quart{30.65}{6.60}{33.28}{100.00} \\
\texttt{NN\_scaled} & 17 & 671.84 & 5.32 & \quart{31.75}{5.32}{33.61}{100.00} \\
\texttt{$k$NN\_onehot} & 16 & 1021.07 & 4.49 & \quart{48.97}{4.49}{51.08}{100.00} \\
\texttt{RF\_scaled} & 15 & 735.36 & 4.61 & \quart{35.10}{4.61}{36.79}{100.00} \\
\texttt{DT\_label} & 14 & 739.20 & 6.96 & \quart{34.49}{6.96}{36.98}{100.00} \\
\texttt{KRR\_onehot} & 13 & 910.47 & 4.64 & \quart{43.52}{4.64}{45.55}{100.00} \\
\texttt{DT\_onehot} & 12 & 803.96 & 8.26 & \quart{36.70}{8.26}{40.22}{100.00} \\
\texttt{DT\_scaled} & 11 & 950.32 & 6.04 & \quart{45.36}{6.04}{47.54}{100.00} \\
\texttt{LR\_label} & 10 & 1046.27 & 3.38 & \quart{50.07}{3.38}{52.34}{100.00} \\
\texttt{$k$NN\_scaled} & 9 & 939.14 & 3.79 & \quart{45.49}{3.79}{46.98}{100.00} \\
\texttt{LR\_scaled} & 8 & 1084.04 & 3.01 & \quart{53.02}{3.01}{54.23}{100.00} \\
\texttt{KRR\_label} & 7 & 1110.85 & 3.61 & \quart{53.77}{3.61}{55.57}{100.00} \\
\texttt{KRR\_scaled} & 6 & 1183.43 & 3.39 & \quart{57.17}{3.39}{59.20}{100.00} \\
\texttt{$k$NN\_label} & 5 & 1455.11 & 4.34 & \quart{70.68}{4.34}{72.79}{100.00} \\
\texttt{SVR\_label} & 4 & 1487.67 & 4.93 & \quart{72.13}{4.93}{74.42}{100.00} \\
\texttt{SVR\_onehot} & 3 & 1566.62 & 5.47 & \quart{76.28}{5.47}{78.37}{100.00} \\
\texttt{SVR\_scaled} & 2 & 1714.36 & 5.89 & \quart{83.50}{5.89}{85.76}{100.00} \\
\texttt{LR\_onehot} & 1 & 9999.99 & 17.91 & \quart{90}{17}{99}{100.00} \\
        \end{tabular}& 
        \begin{tabular}{|lcccl}
           \cellcolor[gray]{1}\textbf{Pair} & \cellcolor[gray]{1}\textbf{Total Score} &     
           & & \\
            \hline
        
\cellcolor{red!10}\texttt{NN\_onehot} & \cellcolor{red!10}66 &\cellcolor{red!10} &\cellcolor{red!10} &\cellcolor{red!10} \hspace{25 mm} \\
\texttt{NN\_label} & 63 & & & \\
\texttt{NN\_scaled} & 63 & & & \\\hline
\texttt{RF\_onehot}& 52 &&&\\
\cellcolor{red!10}\texttt{RF\_label}&\cellcolor{red!10}57 &\cellcolor{red!10}&\cellcolor{red!10}& \cellcolor{red!10}\\
\texttt{RF\_scaled}& 53 &&&\\\hline
\texttt{DT\_onehot}& 44&&&\\
\cellcolor{red!10}\texttt{DT\_label}& \cellcolor{red!10}47 &\cellcolor{red!10}&\cellcolor{red!10}&\cellcolor{red!10}\\
\texttt{DT\_scaled}& 44 &&&\\\hline
\cellcolor{red!10}\texttt{$k$NN\_onehot}& \cellcolor{red!10}39 &\cellcolor{red!10}&\cellcolor{red!10}&\cellcolor{red!10}\\
\texttt{$k$NN\_label}& 24 &&&\\
\texttt{$k$NN\_scaled}& 34 &&&\\\hline
\texttt{LR\_onehot}& 15 &&&\\
\cellcolor{red!10}\texttt{LR\_label}&\cellcolor{red!10}25 &\cellcolor{red!10}&\cellcolor{red!10}&\cellcolor{red!10}\\
\texttt{LR\_scaled}& 23 &&&\\\hline
\cellcolor{red!10}\texttt{KRR\_onehot}& \cellcolor{red!10}39 &\cellcolor{red!10}&\cellcolor{red!10}&\cellcolor{red!10}\\
\texttt{KRR\_label}& 27 &&&\\
\texttt{KRR\_scaled}& 24 &&&\\\hline
\cellcolor{red!10}\texttt{SVR\_onehot}& \cellcolor{red!10}13 &\cellcolor{red!10}&\cellcolor{red!10}&\cellcolor{red!10}\\
\texttt{SVR\_label}& 11 &&&\\
\texttt{SVR\_scaled}& 8 &&&\\
        \end{tabular}\\
        \\
      \textbf{\small (d). \textsc{ExaStencils}}  & \textbf{\small (e). \textsc{x264}}& \textbf{\small (f). Total Scott-Knott scores over all systems}\\
    \end{tabular}
  \end{adjustbox}
   \end{center}
\end{table*}

\begin{figure}[t]
\centering
\includegraphics[width=\columnwidth]{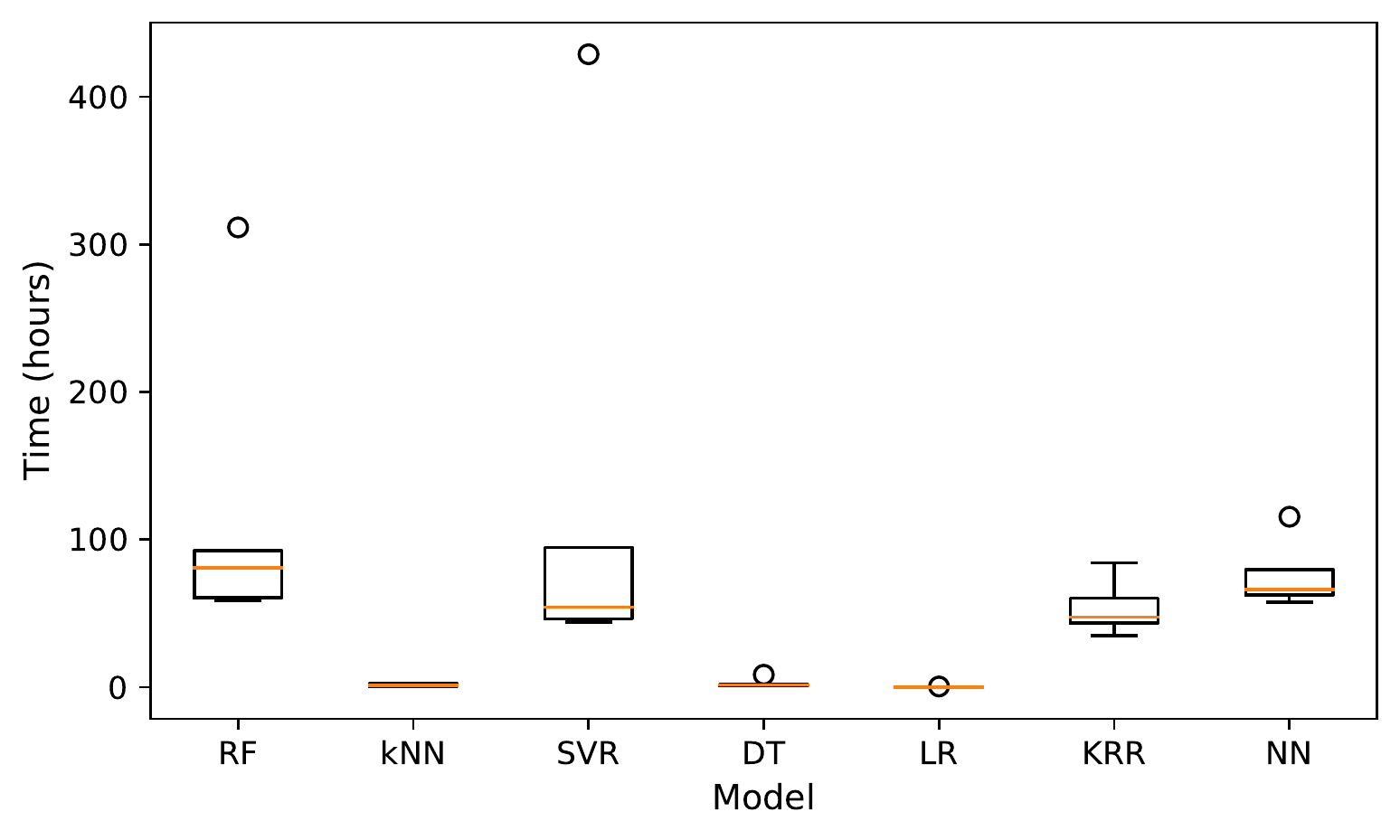}
\caption{The boxplot of the total time required for identifying the best encoding scheme (with respect to a model) over all systems studied.}
\label{fig:total-time}
\end{figure}

\subsection{RQ1: Cost of Trial-and-Error}
\subsubsection{Method}


To answer \textbf{RQ1}, for each encoding scheme, we record the time taken to complete all 50 runs under a model and system (including training, hyperparameter tuning, and evaluation). To identify the best encoding scheme using trial-and-error in a case-by-case manner, the ``efforts'' required would be the total time taken for evaluating a model under all encoding schemes for a system.


\begin{table*}[t!]
  
          \caption{Scott-Knott test, Med (median), and Interquartile Range (IQR) on the training time (minutes) of all models and systems. The format is the same as that for Table~\ref{tb:rq2-plot}.}
\label{tb:rq3-plot}
  \begin{center}
    \begin{adjustbox}{max width = \textwidth}
     
\footnotesize
    \begin{tabular}{c@{}c@{}c}
        \begin{tabular}{lcccl}
            \cellcolor[gray]{1}\textbf{Pair} & \cellcolor[gray]{1}\textbf{Score} &     \cellcolor[gray]{1}\textbf{Med} & \cellcolor[gray]{1}\textbf{IQR} & \cellcolor[gray]{1}\\
            \hline
\texttt{LR\_scaled}  & 19 & <0.01 & 0.00 & \quart{0.00}{0.00}{0.00}{100.00} \\
\texttt{LR\_label}  & 19 & <0.01 & 0.00 & \quart{0.00}{0.00}{0.00}{100.00} \\
\texttt{LR\_onehot}  & 18 & 0.02 & 0.01 & \quart{0.03}{0.01}{0.03}{100.00} \\
\texttt{DT\_label}  & 17 & 0.49 & 0.01 & \quart{0.75}{0.01}{0.76}{100.00} \\
\texttt{DT\_scaled}  & 16 & 0.52 & 0.01 & \quart{0.80}{0.01}{0.81}{100.00} \\
\texttt{$k$NN\_label} & 15 & 0.84 & 0.03 & \quart{1.29}{0.03}{1.31}{100.00} \\
\texttt{$k$NN\_onehot} & 14 & 0.86 & 0.02 & \quart{1.33}{0.02}{1.34}{100.00} \\
\texttt{DT\_onehot}  & 13 & 1.07 & 0.04 & \quart{1.65}{0.04}{1.67}{100.00} \\
\texttt{$k$NN\_scaled}  & 12 & 1.36 & 0.05 & \quart{2.09}{0.05}{2.12}{100.00} \\
\texttt{NN\_scaled} & 11 & 21.61 & 1.30 & \quart{33.55}{1.30}{33.74}{100.00} \\
\texttt{NN\_onehot} & 10 & 23.76 & 0.20 & \quart{36.99}{0.20}{37.09}{100.00} \\
\texttt{RF\_label} & 9 & 28.71 & 0.33 & \quart{44.70}{0.33}{44.83}{100.00} \\
\texttt{KRR\_scaled}  & 8 & 28.86 & 0.08 & \quart{45.02}{0.08}{45.06}{100.00} \\
\texttt{NN\_label} & 7 & 29.02 & 0.24 & \quart{45.23}{0.24}{45.31}{100.00} \\
\texttt{KRR\_onehot}  & 6 & 29.82 & 0.08 & \quart{46.52}{0.08}{46.57}{100.00} \\
\texttt{SVR\_label}  & 5 & 30.45 & 0.54 & \quart{47.11}{0.54}{47.55}{100.00} \\
\texttt{RF\_scaled}  & 4 & 34.20 & 0.62 & \quart{53.06}{0.62}{53.41}{100.00} \\
\texttt{SVR\_onehot} & 3 & 34.75 & 1.61 & \quart{53.53}{1.61}{54.26}{100.00} \\
\texttt{KRR\_label}  & 2 & 42.28 & 0.18 & \quart{65.94}{0.18}{66.02}{100.00} \\
\texttt{RF\_onehot} & 1 & 47.84 & 1.62 & \quart{73.81}{1.62}{74.69}{100.00} \\
\texttt{SVR\_scaled}  & 1 & 60.96 & 47.97 & \quart{48.66}{47.97}{95.18}{100.00} \\
        \end{tabular} & 
        \begin{tabular}{|lcccl|}
           \cellcolor[gray]{1}\textbf{Pair} & \cellcolor[gray]{1}\textbf{Score} &     \cellcolor[gray]{1}\textbf{Med} & \cellcolor[gray]{1}\textbf{IQR} & \cellcolor[gray]{1}\\
            \hline
\texttt{LR\_scaled}  & 20 & <0.01 & 0.00 & \quart{0.00}{0.00}{0.00}{100.00} \\
\texttt{LR\_label}  & 19 & <0.01 & 0.00 & \quart{0.00}{0.00}{0.00}{100.00} \\
\texttt{LR\_onehot}  & 18 & 0.01 & 0.01 & \quart{0.01}{0.01}{0.01}{100.00} \\
\texttt{$k$NN\_label} & 17 & 0.23 & 0.00 & \quart{0.40}{0.00}{0.40}{100.00} \\
\texttt{$k$NN\_scaled}  & 16 & 0.30 & 0.06 & \quart{0.49}{0.06}{0.52}{100.00} \\
\texttt{DT\_scaled}  & 15 & 0.30 & 0.01 & \quart{0.52}{0.01}{0.53}{100.00} \\
\texttt{DT\_label}  & 14 & 0.31 & 0.01 & \quart{0.54}{0.01}{0.55}{100.00} \\
\texttt{$k$NN\_onehot} & 13 & 0.80 & 0.03 & \quart{1.39}{0.03}{1.41}{100.00} \\
\texttt{DT\_onehot}  & 12 & 0.85 & 0.01 & \quart{1.50}{0.01}{1.51}{100.00} \\
\texttt{KRR\_label}  & 11 & 14.72 & 0.57 & \quart{25.85}{0.57}{26.03}{100.00} \\
\texttt{KRR\_scaled}  & 10 & 14.92 & 0.03 & \quart{26.36}{0.03}{26.37}{100.00} \\
\texttt{RF\_scaled}  & 9 & 17.01 & 0.48 & \quart{29.85}{0.48}{30.07}{100.00} \\
\texttt{RF\_label} & 8 & 17.15 & 0.38 & \quart{30.18}{0.38}{30.31}{100.00} \\
\texttt{SVR\_label}  & 7 & 17.73 & 0.58 & \quart{31.21}{0.58}{31.34}{100.00} \\
\texttt{SVR\_scaled}  & 6 & 18.69 & 0.35 & \quart{32.80}{0.35}{33.03}{100.00} \\
\texttt{KRR\_onehot}  & 5 & 22.22 & 1.09 & \quart{39.20}{1.09}{39.28}{100.00} \\
\texttt{SVR\_onehot} & 4 & 28.41 & 0.98 & \quart{49.43}{0.98}{50.22}{100.00} \\
\texttt{NN\_scaled} & 3 & 27.85 & 4.20 & \quart{49.03}{4.20}{49.23}{100.00} \\
\texttt{NN\_label} & 2 & 28.03 & 8.81 & \quart{47.80}{8.81}{49.54}{100.00} \\
\texttt{NN\_onehot} & 1 & 35.36 & 5.16 & \quart{60.76}{5.16}{62.51}{100.00} \\
\texttt{RF\_onehot} & 1 & 36.77 & 0.50 & \quart{64.82}{0.50}{65.00}{100.00} \\
        \end{tabular} &
                \begin{tabular}{lcccl}
             \cellcolor[gray]{1}\textbf{Pair} & \cellcolor[gray]{1}\textbf{Score} &     \cellcolor[gray]{1}\textbf{Med} & \cellcolor[gray]{1}\textbf{IQR} & \cellcolor[gray]{1}
            \\
            \hline
\texttt{LR\_scaled}  & 20 & <0.01 & 0.00 & \quart{0.00}{0.00}{0.00}{100.00} \\
\texttt{LR\_label}  & 20 & <0.01 & 0.00 & \quart{0.00}{0.00}{0.00}{100.00} \\
\texttt{LR\_onehot}  & 19 & 0.01 & 0.01 & \quart{0.01}{0.01}{0.01}{100.00} \\
\texttt{$k$NN\_label} & 18 & 0.17 & 0.00 & \quart{0.26}{0.00}{0.26}{100.00} \\
\texttt{$k$NN\_scaled}  & 17 & 0.25 & 0.02 & \quart{0.37}{0.02}{0.38}{100.00} \\
\texttt{DT\_scaled}  & 16 & 0.26 & 0.00 & \quart{0.39}{0.00}{0.39}{100.00} \\
\texttt{DT\_label}  & 15 & 0.26 & 0.00 & \quart{0.40}{0.00}{0.40}{100.00} \\
\texttt{$k$NN\_onehot} & 14 & 0.47 & 0.00 & \quart{0.72}{0.00}{0.72}{100.00} \\
\texttt{DT\_onehot}  & 13 & 1.36 & 0.03 & \quart{2.06}{0.03}{2.07}{100.00} \\
\texttt{KRR\_label}  & 12 & 13.78 & 0.03 & \quart{20.99}{0.03}{21.00}{100.00} \\
\texttt{KRR\_scaled}  & 11 & 13.75 & 0.05 & \quart{20.95}{0.05}{20.96}{100.00} \\
\texttt{KRR\_onehot}  & 10 & 14.35 & 0.50 & \quart{21.46}{0.50}{21.88}{100.00} \\
\texttt{RF\_scaled}  & 9 & 16.75 & 0.15 & \quart{25.45}{0.15}{25.53}{100.00} \\
\texttt{RF\_label} & 8 & 16.90 & 0.27 & \quart{25.63}{0.27}{25.76}{100.00} \\
\texttt{SVR\_scaled}  & 7 & 17.11 & 0.04 & \quart{26.07}{0.04}{26.09}{100.00} \\
\texttt{SVR\_label}  & 6 & 17.33 & 0.14 & \quart{26.37}{0.14}{26.41}{100.00} \\
\texttt{SVR\_onehot} & 5 & 21.01 & 0.14 & \quart{32.00}{0.14}{32.03}{100.00} \\
\texttt{NN\_label} & 4 & 22.25 & 9.76 & \quart{29.07}{9.76}{33.92}{100.00} \\
\texttt{NN\_scaled} & 3 & 23.07 & 9.06 & \quart{30.40}{9.06}{35.17}{100.00} \\
\texttt{NN\_onehot} & 2 & 22.13 & 10.26 & \quart{29.52}{10.26}{33.73}{100.00} \\
\texttt{RF\_onehot} & 1 & 63.34 & 1.15 & \quart{96.07}{1.15}{96.55}{100.00} \\
        \end{tabular} \\
        \\
         \textbf{\small (a). \textsc{MongoDB}} & \small \textbf{(b). \textsc{Lrzip}} & \textbf{\small (c). \textsc{Trimesh}}
        \\
        \\
        \begin{tabular}{lcccl}
            \cellcolor[gray]{1}\textbf{Pair} & \cellcolor[gray]{1}\textbf{Score} &     \cellcolor[gray]{1}\textbf{Med} & \cellcolor[gray]{1}\textbf{IQR} & \cellcolor[gray]{1}\\
            \hline
\texttt{LR\_scaled}  & 20 & <0.01 & 0.00 & \quart{0.00}{0.00}{0.00}{100.00} \\
\texttt{LR\_label}  & 19 & <0.01 & 0.00 & \quart{0.00}{0.00}{0.00}{100.00} \\
\texttt{LR\_onehot}  & 18 & 0.01 & 0.00 & \quart{0.01}{0.00}{0.01}{100.00} \\
\texttt{$k$NN\_label} & 17 & 0.18 & 0.00 & \quart{0.27}{0.00}{0.27}{100.00} \\
\texttt{$k$NN\_scaled}  & 16 & 0.22 & 0.02 & \quart{0.33}{0.02}{0.34}{100.00} \\
\texttt{DT\_scaled}  & 15 & 0.23 & 0.00 & \quart{0.35}{0.00}{0.35}{100.00} \\
\texttt{DT\_label}  & 14 & 0.26 & 0.01 & \quart{0.39}{0.01}{0.40}{100.00} \\
\texttt{$k$NN\_onehot} & 13 & 0.49 & 0.05 & \quart{0.71}{0.05}{0.75}{100.00} \\
\texttt{DT\_onehot}  & 12 & 0.73 & 0.01 & \quart{1.10}{0.01}{1.11}{100.00} \\
\texttt{KRR\_scaled}  & 11 & 13.96 & 0.15 & \quart{21.22}{0.15}{21.24}{100.00} \\
\texttt{KRR\_onehot}  & 10 & 15.30 & 2.39 & \quart{22.35}{2.39}{23.29}{100.00} \\
\texttt{RF\_label} & 9 & 15.64 & 0.35 & \quart{23.59}{0.35}{23.80}{100.00} \\
\texttt{RF\_scaled}  & 8 & 15.86 & 0.26 & \quart{23.93}{0.26}{24.13}{100.00} \\
\texttt{SVR\_scaled}  & 7 & 16.98 & 0.10 & \quart{25.78}{0.10}{25.84}{100.00} \\
\texttt{SVR\_label}  & 6 & 16.48 & 3.21 & \quart{25.03}{3.21}{25.07}{100.00} \\
\texttt{NN\_onehot} & 5 & 18.02 & 4.46 & \quart{25.91}{4.46}{27.42}{100.00} \\
\texttt{SVR\_onehot} & 5 & 18.50 & 0.07 & \quart{28.12}{0.07}{28.15}{100.00} \\
\texttt{KRR\_label}  & 4 & 21.01 & 21.64 & \quart{29.63}{21.64}{31.97}{100.00} \\
\texttt{NN\_scaled} & 3 & 25.70 & 19.07 & \quart{35.98}{19.07}{39.10}{100.00} \\
\texttt{NN\_label} & 2 & 29.11 & 24.35 & \quart{34.46}{24.35}{44.29}{100.00} \\
\texttt{RF\_onehot} & 1 & 32.51 & 44.27 & \quart{49.23}{44.27}{49.47}{100.00} \\
        \end{tabular} & 
        \begin{tabular}{|lcccl}
           \cellcolor[gray]{1}\textbf{Pair} & \cellcolor[gray]{1}\textbf{Score} &     \cellcolor[gray]{1}\textbf{Med} & \cellcolor[gray]{1}\textbf{IQR} & \cellcolor[gray]{1}\\
            \hline
\texttt{LR\_label}  & 19 & <0.01 & 0.00 & \quart{0.00}{0.00}{0.00}{100.00} \\
\texttt{LR\_scaled}  & 18 & <0.01 & 0.00 & \quart{0.00}{0.00}{0.00}{100.00} \\
\texttt{DT\_label}  & 17 & 0.41 & 0.00 & \quart{0.07}{0.00}{0.07}{100.00} \\
\texttt{$k$NN\_label} & 16 & 0.46 & 0.00 & \quart{0.08}{0.00}{0.08}{100.00} \\
\texttt{DT\_scaled}  & 15 & 0.48 & 0.00 & \quart{0.08}{0.00}{0.08}{100.00} \\
\texttt{$k$NN\_scaled}  & 14 & 0.48 & 0.01 & \quart{0.08}{0.01}{0.08}{100.00} \\
\texttt{LR\_onehot}  & 13 & 0.57 & 0.02 & \quart{0.09}{0.02}{0.10}{100.00} \\
\texttt{$k$NN\_onehot} & 12 & 1.98 & 0.04 & \quart{0.31}{0.04}{0.34}{100.00} \\
\texttt{DT\_onehot}  & 11 & 9.17 & 0.05 & \quart{1.53}{0.05}{1.56}{100.00} \\
\texttt{KRR\_label}  & 10 & 18.87 & 0.88 & \quart{2.42}{0.88}{3.20}{100.00} \\
\texttt{SVR\_scaled}  & 9 & 19.69 & 0.42 & \quart{2.96}{0.42}{3.34}{100.00} \\
\texttt{NN\_scaled} & 8 & 16.83 & 0.82 & \quart{2.77}{0.82}{2.86}{100.00} \\
\texttt{KRR\_onehot}  & 7 & 23.76 & 0.21 & \quart{3.98}{0.21}{4.03}{100.00} \\
\texttt{KRR\_scaled}  & 6 & 33.79 & 0.33 & \quart{5.54}{0.33}{5.73}{100.00} \\
\texttt{NN\_label} & 5 & 29.38 & 2.17 & \quart{4.00}{2.17}{4.99}{100.00} \\
\texttt{RF\_scaled}  & 5 & 30.28 & 0.07 & \quart{5.10}{0.07}{5.14}{100.00} \\
\texttt{SVR\_label}  & 5 & 35.82 & 0.15 & \quart{5.98}{0.15}{6.08}{100.00} \\
\texttt{RF\_label} & 4 & 34.27 & 3.76 & \quart{5.64}{3.76}{5.82}{100.00} \\
\texttt{NN\_onehot} & 3 & 81.93 & 4.25 & \quart{11.38}{4.25}{13.90}{100.00} \\
\texttt{RF\_onehot} & 2 & 244.98 & 10.31 & \quart{40.42}{10.31}{41.57}{100.00} \\
\texttt{SVR\_onehot} & 1 & 519.99 & 23.35 & \quart{68.66}{23.35}{88.23}{100.00} \\
        \end{tabular}& 
        \begin{tabular}{|lcccl}
           \cellcolor[gray]{1}\textbf{Pair} & \cellcolor[gray]{1}\textbf{Score} &     
           & & \\
            \hline
\texttt{LR\_onehot}& 86 &&&\\
\texttt{LR\_label}& 96 &&&\\
\cellcolor{red!10}\texttt{LR\_scaled}& \cellcolor{red!10}97 &\cellcolor{red!10}&\cellcolor{red!10}&\cellcolor{red!10}\\
\hline
\texttt{$k$NN\_onehot}& 66 &&&\\
\cellcolor{red!10}\texttt{$k$NN\_label}& \cellcolor{red!10}83 &\cellcolor{red!10}&\cellcolor{red!10}&\cellcolor{red!10}\\
\texttt{$k$NN\_scaled}& 75 &&&\\
\hline
\texttt{DT\_onehot}& 61 &&&\\
\cellcolor{red!10}\texttt{DT\_label}& \cellcolor{red!10}77 &\cellcolor{red!10}&\cellcolor{red!10}&\cellcolor{red!10}\\
\cellcolor{red!10}\texttt{DT\_scaled}& \cellcolor{red!10}77 &\cellcolor{red!10}&\cellcolor{red!10}&\cellcolor{red!10}\\
\hline
\texttt{KRR\_onehot}& 38 &&&\\
\texttt{KRR\_label}& 39 &&&\\
\cellcolor{red!10}\texttt{KRR\_scaled}& \cellcolor{red!10}46 &\cellcolor{red!10}&\cellcolor{red!10}&\cellcolor{red!10}\\
\hline
\texttt{SVR\_onehot}& 18 &&&\\
\texttt{SVR\_label}& 29 &&&\\
\cellcolor{red!10}\texttt{SVR\_scaled}& \cellcolor{red!10}30 &\cellcolor{red!10}&\cellcolor{red!10}&\cellcolor{red!10}\\
\hline
\texttt{RF\_onehot}& 6 &&&\\
\cellcolor{red!10}\texttt{RF\_label}& \cellcolor{red!10}38 &\cellcolor{red!10}&\cellcolor{red!10}&\cellcolor{red!10}\\
\texttt{RF\_scaled}& 35 &&&\\
\hline
\texttt{NN\_onehot} & 21 & & & \hspace{30 mm} \\
\texttt{NN\_label} & 20 & & & \\
\cellcolor{red!10}\texttt{NN\_scaled} & \cellcolor{red!10}28 & \cellcolor{red!10}&\cellcolor{red!10} &\cellcolor{red!10} \\






        \end{tabular}\\
        \\
      \textbf{\small (d). \textsc{ExaStencils}}  & \textbf{\small (e). \textsc{x264}}& \textbf{\small (f). Total Scott-Knott scores over all systems}\\
    \end{tabular}
  \end{adjustbox}
   \end{center}
\end{table*}

\subsubsection{Results}

Figure~\ref{fig:total-time} shows the result, from which we obtain some clear evidence:

\begin{itemize}
    \item \textbf{Finding 1:} It can take an extremely long time to conclude which encoding scheme is better depending on the models: this is almost 100 hours (median) for RF and around 80 hours (median) for SVR in general; it can go up to 400$+$ hours on some systems. For KRR and NN, which takes less time to do so, still requires around at least two and a half days (60$+$ hours).
    \item \textbf{Finding 2:} For certain models, it may be possible to find the best encoding scheme. For example, it takes less than an hour for $k$NN, DT, and LR due to their low computational needs. Yet, whether one would be willing to spend valuable development time for this is really case-dependent.
\end{itemize}

The above confirms that finding the best encoding scheme for learning software performance can be non-trivial and the needs of our study. Therefore, for \textbf{RQ1}, we say:

\begin{quotebox}
   \noindent
   \textit{\textbf{RQ1:} Depending on the model, finding the best encoding scheme using trial-and-error can be highly expensive, as it may take 60$+$ hours (median) and up to 400$+$ hours. However, for other cases, the ``effort" only need less than one hour, which may be acceptable depending on the scenario.}
\end{quotebox}



\subsection{RQ2: Accuracy}
\label{sec:rq2-acc}

  

\begin{figure*}[t]
\centering

\begin{subfigure}{.33\textwidth}
  \centering
  \includegraphics[width=\linewidth]{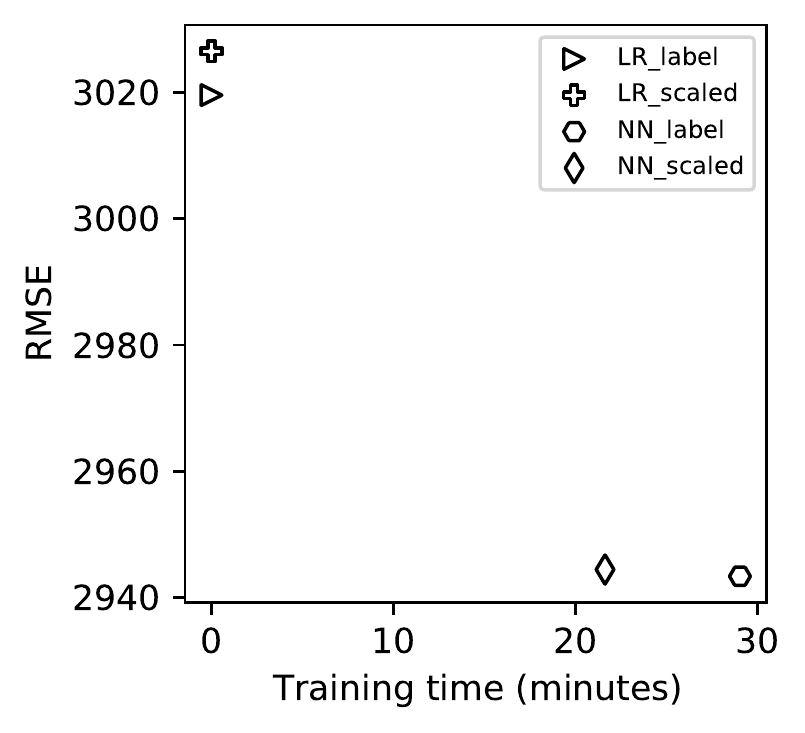}  
  \caption{\textsc{MongoDB}}
  \label{fig:tradeoff-a}
\end{subfigure}
\begin{subfigure}{.33\textwidth}
  \centering
  \includegraphics[width=\linewidth]{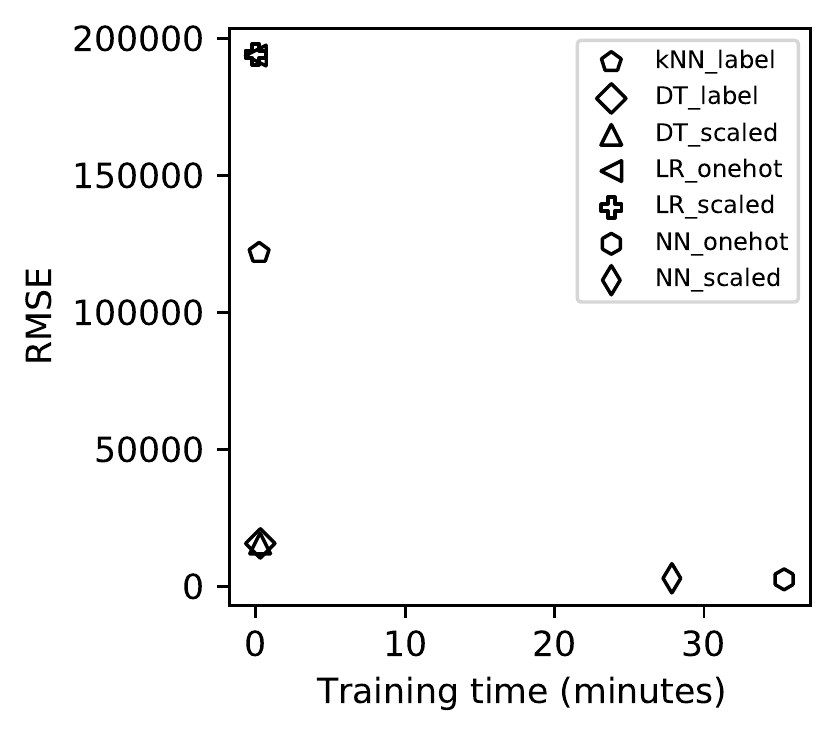} 
  \caption{\textsc{Lrzip}}
  \label{fig:tradeoff-b}
\end{subfigure}
\begin{subfigure}{.33\textwidth}
  \centering
  \includegraphics[width=\linewidth]{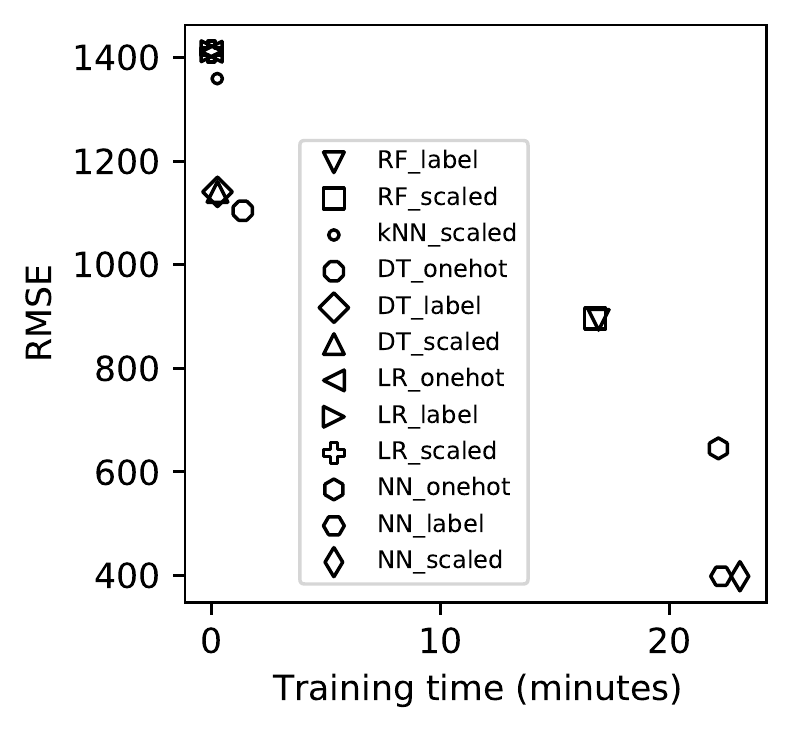} 
  \caption{\textsc{Trimesh}}
  \label{fig:tradeoff-c}
\end{subfigure}


\begin{subfigure}{.33\textwidth}
  \centering
  \includegraphics[width=\linewidth]{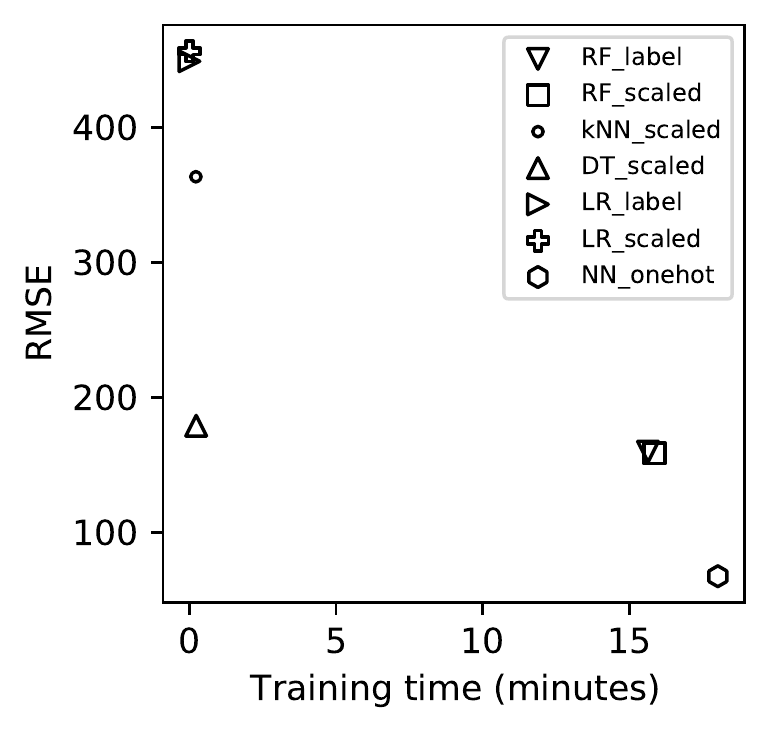}  
  \caption{\textsc{ExaStencils}}
  \label{fig:tradeoff-d}
\end{subfigure}
\begin{subfigure}{.34\textwidth}
  \centering
  \includegraphics[width=\linewidth]{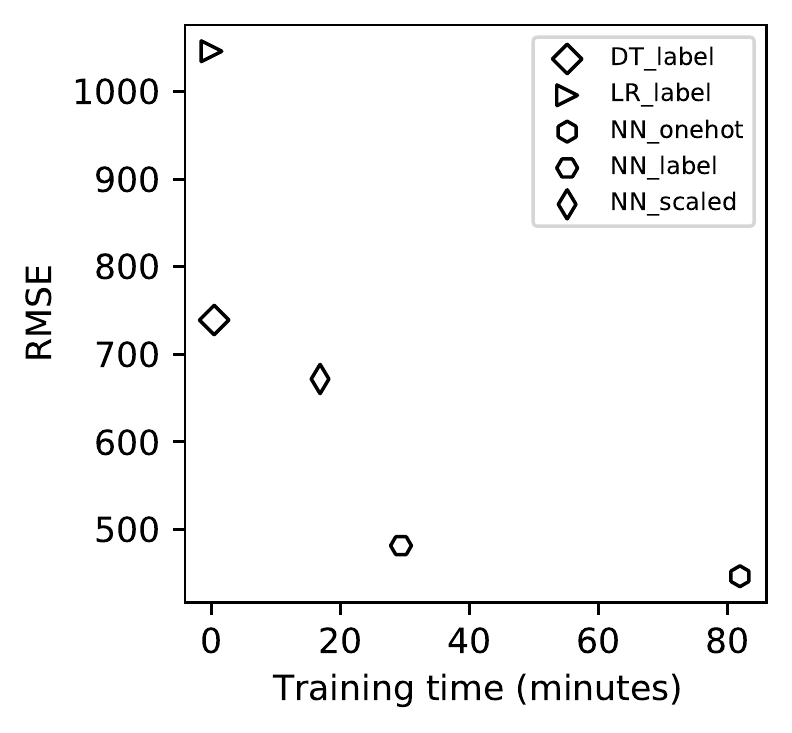} 
  \caption{\textsc{x264}}
  \label{fig:tradeoff-e}
\end{subfigure}

    \caption{The trade-off between RMSE and training time over all Pareto-optimal model-encoding pairs.}
    \label{fig:tradeoff}
\end{figure*}

\subsubsection{Method}
To study \textbf{RQ2}, we compare all RMSE values for the three encoding schemes under the models and systems. That is, for each subject system, there are $3 \times 7 = 21$ pairs of model-encoding (50 RMSE repeats each). To ensure statistical significance among the comparisons, we use Scott-Knott test to assign a score for each pair, hence similar ones are clustered together (same score) while different ones can be ranked (the higher score, the better).


\subsubsection{Results}

As illustrated in Table~\ref{tb:rq2-plot}, we observe some interesting findings:

\begin{itemize}
    \item \textbf{Finding 3:} From Table~\ref{tb:rq2-plot}f, overall, label and one-hot encoding are clearly more accurate than scaled label encoding across the models, as the former two have the best total Scott-Knott scores for all models over the systems studied. Between these two, one-hot encoding tends to be slightly better across all models, as it wins on 4 models against the 3 wins by label encoding. We observe similar trend from Table~\ref{tb:rq2-plot}a to~\ref{tb:rq2-plot}e over the systems.
    \item \textbf{Finding 4:} For different models in Table~\ref{tb:rq2-plot}f, we observe that one-hot encoding is the best for deep learning, lazy, and kernel models while label encoding is preferred on linear and tree models. The same has also been registered in Table~\ref{tb:rq2-plot}a to~\ref{tb:rq2-plot}e.
    \item \textbf{Finding 5:} From Table~\ref{tb:rq2-plot}a to~\ref{tb:rq2-plot}e, NN is clearly amongst the top models on Scott-Knott score and RMSE regardless of the encoding schemes and systems. In particular, when NN is chosen, \texttt{NN\_onehot} is the best, as it has a better Scott-Knott score than the other two on 3 out of 5 systems, draw on one system and lose on the remaining one, leading to a 75\% cases of no worse outcome than \texttt{NN\_label} and \texttt{NN\_scaled}.  
\end{itemize}

To conclude, we can answer \textbf{RQ2} as:

\begin{quotebox}
   \noindent
   \textit{\textbf{RQ2:} In general, the one-hot encoding tends to have the best accuracy and the scaled label encoding should be avoided. In particular, \texttt{NN\_onehot} is the safest option for the overall optimal accuracy among the subjects studied.}
\end{quotebox}

\subsection{RQ3: Training Time}
\label{sec:rq3-time}


\subsubsection{Research}
Similar to \textbf{RQ2}, here we measured the training time over 50 runs for all 21 pairs of model-encoding for each system. 

\subsubsection{Results}

With Table~\ref{tb:rq3-plot}, we can observe some patterns:

\begin{itemize}
    \item \textbf{Finding 6:} Overall, from Table~\ref{tb:rq3-plot}f, label and scaled label encoding are much faster to train than their one-hot counterpart, which has never won the other two under any model across the systems. In particular, scaled label encoding appears to have the fastest training than others in general, as the former wins on 4 models, draws on one, and loses only on two. Similar results have been obtained in Table~\ref{tb:rq3-plot}a to~\ref{tb:rq3-plot}e.
    
    
    \item \textbf{Finding 7:} For different model types, in Table~\ref{tb:rq3-plot}f, the label encoding tends to be the best option in terms of training time for tree model and lazy model; the scaled label counterpart is faster on deep learning model, linear model, and kernel model. This is similar to that from Table~\ref{tb:rq3-plot}a to~\ref{tb:rq3-plot}e.
    
    \item \textbf{Finding 8:} From Table~\ref{tb:rq3-plot}a to~\ref{tb:rq3-plot}e, unexpectedly LR has the fastest training time over all systems and this model works the best with scaled label encoding since \texttt{LR\_scaled} is the fastest on 4 out of 5 systems; its difference to \texttt{LR\_label} tends to be marginal though. 
\end{itemize}

Therefore, we say:

\begin{quotebox}
   \noindent
   \textit{\textbf{RQ3:} The scaled label encoding tends to have the fastest training while one-hot encoding takes the longest time to train. In particular, \texttt{LR\_scaled} is the best choice for the overall fastest training time over the subjects studied.}
\end{quotebox}


\subsection{RQ4: Trade-off Analysis}
\label{sec:rq4-tradeoff}

\subsubsection{Method}

\begin{figure}[t]
\centering

\begin{subfigure}{.31\columnwidth}
  \centering
  \includegraphics[width=\linewidth]{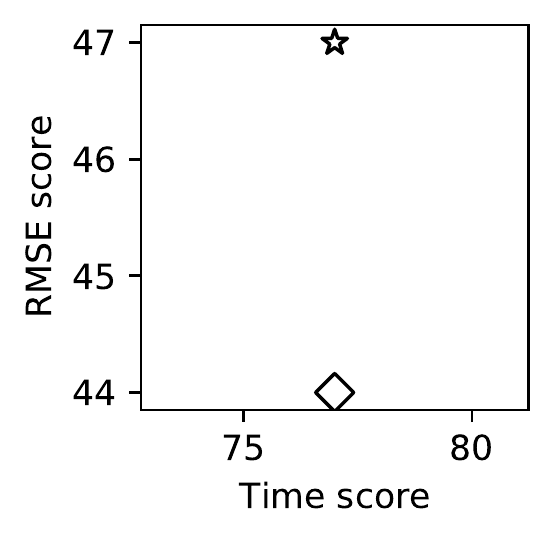}  
  \caption{\texttt{DT}}
  \label{fig:tradeoff-a}
\end{subfigure}
\begin{subfigure}{.31\columnwidth}
  \centering
  \includegraphics[width=\linewidth]{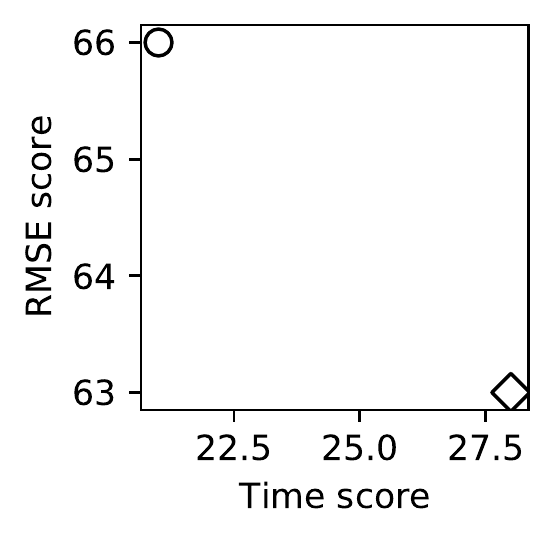} 
  \caption{\texttt{NN}}
  \label{fig:tradeoff-b}
\end{subfigure}
\begin{subfigure}{.31\columnwidth}
  \centering
  \includegraphics[width=\linewidth]{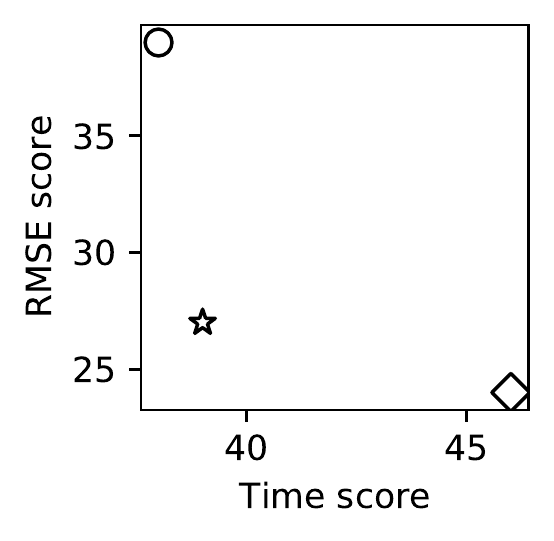} 
  \caption{\texttt{KRR}}
  \label{fig:tradeoff-c}
\end{subfigure}


\begin{subfigure}{.31\columnwidth}
  \centering
  \includegraphics[width=\linewidth]{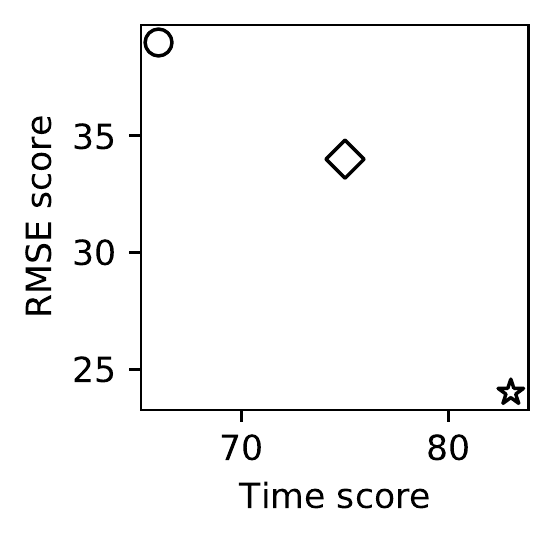}  
  \caption{\texttt{$k$NN}}
  \label{fig:tradeoff-d}
\end{subfigure}
\begin{subfigure}{.325\columnwidth}
  \centering
  \includegraphics[width=\linewidth]{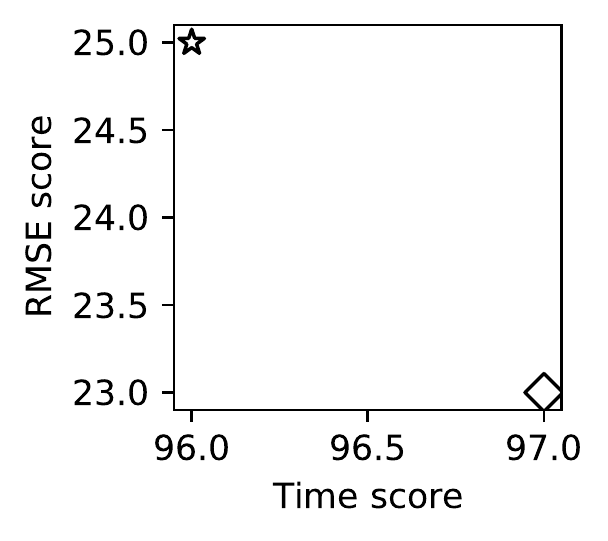} 
  \caption{\texttt{LR}}
  \label{fig:tradeoff-e}
\end{subfigure}
\begin{subfigure}{.31\columnwidth}
  \centering
  \includegraphics[width=\linewidth]{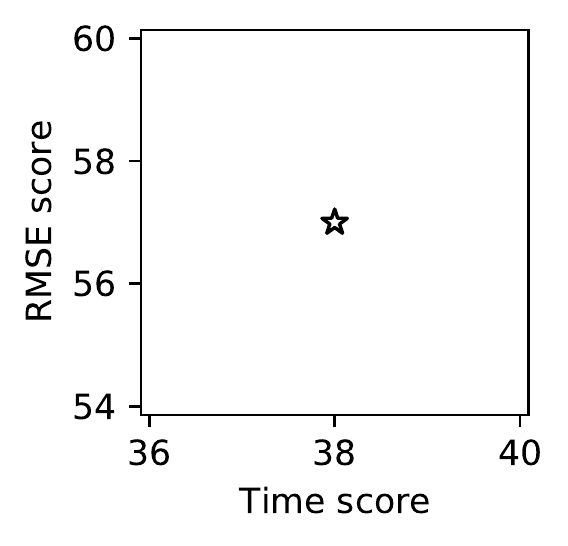}  
  \caption{\texttt{RF}}
  \label{fig:tradeoff-d}
\end{subfigure}

\begin{subfigure}{.31\columnwidth}
  \centering
  \includegraphics[width=\linewidth]{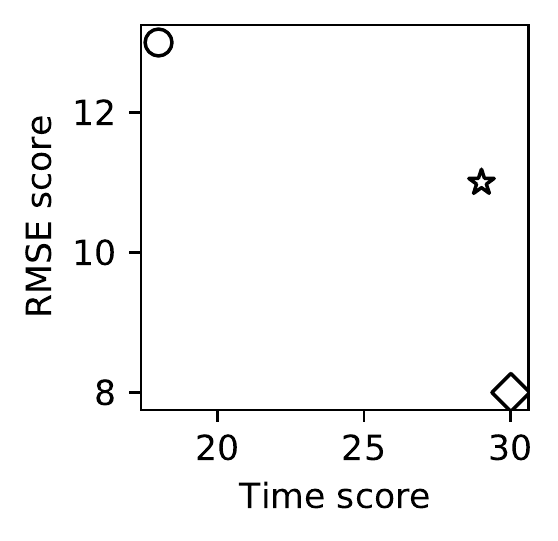} 
  \caption{\texttt{SVR}}
  \label{fig:tradeoff-e}
\end{subfigure}
\begin{subfigure}{.31\columnwidth}
  \centering
  \includegraphics[width=\linewidth]{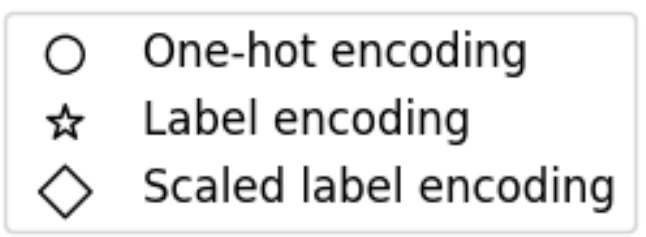} 
  \label{fig:tradeoff-f}
  \vspace{8mm}
\end{subfigure}

    \caption{The total Scott-Knott scores for each model over all systems. Only the Pareto-optimal choice are presented.}
    \label{fig:tradeoff1}
\end{figure}

Understanding \textbf{RQ4} requires us to simultaneously consider the accuracy and training time achieved by all 21 pairs of model-encoding. According to the guidance provided by Li and Chen~\cite{Li2020}, for each system, we seek to analyze the Pareto optimal choices as those are the ones that require trade-offs. Suppose that a pair $P_x$ has $\{A_x, T_x\}$ and another $P_y$ comes with $\{A_y, T_y\}$, whereby $A_x$ and $A_y$ are their median RMSE (over 50 runs) while $T_x$ and $T_y$ are their median training time, respectively. We say $P_x$ dominates $P_y$ if $A_x \leq A_y$ and $T_x \leq T_y$ while there is either $A_x < A_y$ or $T_x < T_y$. A pair, which is not dominated by any other pairs from the total set of 21, is called a Pareto-optimal pair therein. The set of all Pareto-optimal points is called the Pareto front (Figures~\ref{fig:tradeoff}). We also plot the Pareto front with respect to the total Scott-Knott scores (over all systems) under each model (Figures~\ref{fig:tradeoff1}).

Here, a Pareto-optimal pair that has the best accuracy or the fastest training time is called a biased point (or an extreme point). Among others, we are interested in the non-extreme, less biased points, especially those with a well-balanced trade-off.


\subsubsection{Results}

The results are illustrated in Figures~\ref{fig:tradeoff} and \ref{fig:tradeoff1}, from which we obtain some interesting observations:

\begin{itemize}

    \item \textbf{Finding 9:} Over all the model-encoding pairs (Figures~\ref{fig:tradeoff}), label and scaled label encoding can more commonly lead to less biased results (non-extreme points) in the Pareto front than their one-hot counterpart. This can clearly offer more trade-off choices.

    \item \textbf{Finding 10:} In Figures~\ref{fig:tradeoff}, the scaled label encoding tends to achieve more balanced trade-off than the others, but the paired model may vary, i.e., it can be NN, DT, or RF, across the systems. For example, it is \texttt{NN\_scaled} on \textsc{MongoDB} but becomes \texttt{DT\_scaled} on \textsc{Lrzip}.
    
    \item \textbf{Finding 11:} Kernel models like KRR and SVR have never produced Pareto-optimal outcomes over the 5 systems studied (Figures~\ref{fig:tradeoff}).
    
    \item \textbf{Finding 12:} From Figure~\ref{fig:tradeoff1}, only kernel and lazy models have relatively less biased point in their own Pareto front, which are achieved by label (Figure~\ref{fig:tradeoff1}c and~\ref{fig:tradeoff1}g) and scaled label encoding (Figure~\ref{fig:tradeoff1}d), respectively.
    
    
\end{itemize}

In summary, we have:

\begin{quotebox}
   \noindent
   \textit{\textbf{RQ4:} For all the model-encoding pairs, label and scaled label encoding tend to be less biased to accuracy or training time than their one-hot counterpart. In particular, scaled label encoding can lead to relatively more balanced outcomes. However, the paired model for the above may differ depending on the system, but it would never be KRR or SVR, which produce no Pareto-optimal pairs.}
\end{quotebox}




\section{Actionable Suggestions}
\label{sec: discussions}

In this section, we discuss the suggestions on the encoding scheme for learning software performance under a variety of circumstances.





\begin{insightbox}
   \raisebox{-0.05ex}{}
   \tcblower
   \textit{\textbf{Suggestion 1:} When RF, SVR, KRR or NN is to be used, we do not recommend trial-and-error to find the best encoding scheme. However, for $k$NN, DT or LR, it may be practical to ``try them all''.}
\end{insightbox}

From \textbf{RQ1}, it can be rather time-consuming for comparing all three encoding schemes under RF, SVR, KRR or NN. Indeed, the ``efforts'' may be reduced if we consider, e.g., less repeated runs or even reduced data samples. However, to provide a reliable choice, what we consider in this study is essential, and hence further reducing them may increase the instability of the result. The process can be even more expensive if different models are also to be assessed during the trial-and-error. In contrast, when $k$NN, DT, or LR is to be used, it only requires less than one hour each --- an assumption that may be more acceptable within the development lifecycle.


\begin{insightbox}
   \raisebox{-0.05ex}{}
   \tcblower
   \textit{\textbf{Suggestion 2:} 
   When the accuracy is all that matters, among all possible models studied, we recommend using NN paired with one-hot encoding. When a certain model needs to be used, we suggest avoiding scaled label encoding in general and following one-hot encoding for deep learning and kernel models; label encoding for linear and tree models.}
\end{insightbox}

Reflecting on \textbf{RQ2}, when only the accuracy is of concern, we can make suggestions for practitioners to infer the best choice of encoding schemes when experimental assessment is not possible or desirable. Among others, it is clear that NN tends to offer the best accuracy, and NN paired with one-hot encoding, i.e., \texttt{NN\_onehot}, is the most reliable choice. In contrast, scaled label encoding often performs the worst, and hence scaled label encoding can be ruled out from the suggestions.

Besides the fact that the one-hot encoding can generally lead to the best accuracy over the models, we do observe some specific patterns when the model to be used is fixed: one-hot encoding for deep learning and kernel models while label encoding for linear and tree models. 




\begin{insightbox}
   \raisebox{-0.05ex}{}
   \tcblower
   \textit{\textbf{Suggestion 3:} When faster training time is more preferred (e.g., the model needs to be rapidly retrained at runtime), over all models studied, we recommend using linear regression paired with scaled label encoding. When the model is fixed, we suggest adopting scaled label encoding in general (especially for deep learning, linear, and kernel models) and label encoding for tree and lazy models; one-hot encoding should be avoided.}
\end{insightbox}

Deriving from the findings for \textbf{RQ3}, if the training time is of higher importance, we can also estimate the suitable choice of encoding scheme in the absence of experimental evaluation. Over all possible models studied, linear regression is unexpectedly the fastest to train and when it is paired with scaled label encoding (\texttt{LR\_scaled}) the training is the fastest. One-hot encoding is often the slowest to train, and hence can be avoided. 

Although the scaled label encoding appears to be faster to train than its label counterpart, they remain competitive. In fact, when the model to be used has been pre-defined, we observe some common patterns: the scaled label encoding is the best for deep learning, linear, and kernel models while the label encoding is preferred for tree and lazy models.



\begin{insightbox}
   \raisebox{-0.05ex}{}
   \tcblower
   \textit{\textbf{Suggestion 4:} When the preference between accuracy and training time is unclear while the unbiased outcome is preferred, over all models studied, we recommend using scaled label encoding, but the paired model needs some efforts to determine. We certainly suggest avoiding one-hot encoding and kernel model (KRR and SVR regardless of its encoding schemes). When the model is fixed to the kernel and lazy models, the label and scaled label encoding can be chosen to reduce the bias, respectively.
   }
\end{insightbox}

It is not uncommon that the preference between accuracy and training time can be unclear, and hence an unbiased outcome is important. According to the findings for \textbf{RQ4}, this needs the label and scaled label encoding. Because in this case, as we have shown, they often lead to results that are in the middle of the Pareto front for the pairs. In particular, scaled label encoding can often lead to well-balanced results in contrast to the other, but the paired model may vary. We would also suggest avoiding one-hot encoding and kernel model (SVR and KRR), as the former would easily bias to accuracy or training time while the latter leads to no Pareto optimal choice at all over the systems studied.

However, when the model needs to be fixed, only the kernel and lazy models can have less biased choices, which are under the label and scaled label encoding, respectively.

\section{Discussion}
\label{sec:new-discussions}

We now discuss a few interesting points derived from our study.





\subsection{Practicality of Performance Models}

The performance models built can be used in different practical scenarios, under each of which the accuracy and training time can be of great importance (and thereby the choice of encoding schemes are equally crucial).

\subsubsection{Configuration debugging}

Ill-fitted Configurations can lead to bugs such that the resulted performance is dramatically worse than the expectation. Here, a performance model can help software engineers easily inspect which configuration options are likely to be the root cause of the bug and identify the potential fix~\cite{DBLP:conf/sigsoft/XuJFZPT15}. The fact that the model makes inferences without running the system can greatly improve the efficiency of the debugging process. Further, by analyzing the models, software engineers can gain a better understanding of the system's performance characteristics which helps to prevent future configuration bugs.


\subsubsection{Speed up automatic configuration tuning}

Automatic configuration tuning is necessary to optimize the performance of the software system at deployment time. However, due to the expensiveness of measuring the performance, tuning is often a slow and time-consuming process. As one resolution to that issue, the performance model can serve as the surrogate for cheap evaluation of the configuration. Indeed, there have been a few successful applications in this regard, such as those that rely on Bayesian optimization~\cite{DBLP:journals/corr/abs-1801-02175,DBLP:conf/mascots/JamshidiC16}.

\subsubsection{Runtime self-adaptation}

Self-adapting the configuration at runtime is a promising way to manage the system's performance under uncertain environments. In this context, the performance model can help to achieve the adaptation in a timely manner, as it offers a relatively cheap way to reason about the better or worse of different configurations under changing environmental conditions. From the literature of self-adaptive systems, it is not uncommon to see that the performance models are often used during the planning stage~\cite{DBLP:journals/tsc/ChenB17,DBLP:conf/wosp/0001BWY18,DBLP:journals/csur/ChenBY18,DBLP:journals/pieee/ChenBY20,ChenLiDOS22,DBLP:conf/icse/ChenB14}.


\subsection{Why Considering Different Models?}

We note that some learning models perform overwhelmingly better than the others, such as NN. Yet, our study involves a diverse set of models because, in practice, there may be other reasons that a learning model is preferred. For example, linear and tree models may be used as they are directly interpretable~\cite{DBLP:conf/icse/SiegmundKKABRS12,DBLP:conf/splc/ValovGC15}, despite that they can lead to inferior accuracy overall. Therefore, our results on the choice of encoding schemes provide evidence for a wide set of scenarios and the possibility that different models may be involved. 

The other reason for considering different models is that we seek to examine whether the choice of machine learning model matters when deicing what encoding schemes to use. Indeed, our results show that the paired model is an integral part and we provide detailed suggestions in that regard.



\subsection{On Interactions between Configuration Options}

The encoding schemes can serve as different ways to represent the interactions between configuration options. Since the one-hot encoding embeds the values of options as the feature dimensions and captures their interactions, it models a much more finer-grained feature space compared with that of the label and scaled-label counterparts. Our results show that, indeed, such a finer-grained capture of interactions enables one-hot encoding to become the most reliable scheme across the models/software as it has the generally best accuracy. This confirms the current understanding that the interaction between configuration options is important and the way how they are handled can significantly influence the accuracy~\cite{DBLP:conf/icse/SiegmundKKABRS12}. Most importantly, our findings show that it is possible to better handle the interactions at the level of encoding.



\section{Threats to validity}
\label{sec:tov}

Similar to many empirical studies in software engineering, our work is subject to threats to validity. Specifically, \textbf{internal threats} can be related to the configuration options used and their ranges. Indeed, a different set may lead to a different result in some cases. However, here we follow what has been commonly used in state-of-the-art studies, which are representatives for the subject systems. The hyperparameter of the models to tune can also impose this threat. Ideally, widening the set of hyperparameters to tune can complement our results. Yet, considering an extensive set of hyperparameters is rather expensive, as the tuning needs to go through the full training and validation process. To mitigate such, we have examined different hyperparameters in preliminary runs for finding a balance between effectiveness and overhead. 


\textbf{Construct threats} to validity can be related to the metric used. While different metrics exist for measuring accuracy, here we use RMSE, which is a widely used one for learning software performance. The results are also evaluated validated Scott-Knott test~\cite{DBLP:journals/tse/MittasA13}. We also set a data samples of $5,000$, which tends to be reasonable as this is what has been commonly used in prior work~\cite{DBLP:conf/kbse/DornAS20,DBLP:conf/im/JohnssonMS19, DBLP:conf/icdm/ShaoWL19, DBLP:conf/icse/Gerostathopoulos18}. Indeed, using other metrics or different sample size may offer new insights, which we plan to do in future work.

Finally, \textbf{external threats} to validity can raise from the subjects and models used. To mitigate such, we study five commonly studied systems that are of diverse characteristics, together with seven widely-used models. This leads to a total of 105 cases of investigation. Such a setting, although not exhaustive, is not uncommon in empirical software engineering and can serve as a strong foundation to generalize our findings, especially considering that an exhaustive study of all possible models and systems is unrealistic. Yet, we agree that additional subjects may prove fruitful.



\section{Related Work}
\label{sec:related}


A most widely used representation for building machine learning-based software performance model is the one-hot encoding~\cite{DBLP:conf/icse/SiegmundKKABRS12,DBLP:conf/kbse/GuoCASW13,DBLP:conf/kbse/BaoLWF19}. The root motivation of such encoding is derived from the fact that a configurable system can be represented by the feature model --- a tree-liked structure that captures the variability~\cite{DBLP:conf/icse/SiegmundKKABRS12}. In a feature model, each feature can be selected or deselected, which is naturally a binary option. Note that categorical and numeric configuration options can also be captured in the feature model, as long as they can be discretized~\cite{DBLP:journals/tosem/ChenLBY18}. Following this, several approaches have been developed using machine learning. Among others, \citeauthor{DBLP:conf/kbse/GuoCASW13}~\cite{DBLP:conf/kbse/GuoCASW13} use the one-hot encoding combined with the DT to predict software performance, as it fits well with the feature model. \citeauthor{DBLP:conf/kbse/BaoLWF19}~\cite{DBLP:conf/kbse/BaoLWF19} also use the same encoding, and their claim is that it can better capture the options which have no ordinal relationships.

The other, perhaps more natural, encoding scheme for learning software performance is the label encoding, which has also been followed by many studies, either with~\cite{DBLP:conf/icse/Chen19b,DBLP:journals/tse/ChenB17,DBLP:conf/icse/HaZ19} or without scaling~\cite{DBLP:journals/corr/abs-1801-02175,DBLP:conf/sigsoft/SiegmundGAK15,DBLP:conf/sigsoft/NairMSA17}. For example, \citeauthor{DBLP:journals/tse/ChenB17}~\cite{DBLP:journals/tse/ChenB17} directly encode the configuration options to learn the performance model with normalization to $[0,1]$. \citeauthor{DBLP:conf/sigsoft/SiegmundGAK15}~\cite{DBLP:conf/sigsoft/SiegmundGAK15} also follows the label encoding, but the binary and numeric configuration options are treated differently in the model learned with no normalization.

However, the choice between those two encoding schemes for software performance learning often lacks systematic justification, which is the gap that this empirical study aims to bridge.

In the other domains, the importance of choosing the encoding schemes for building machine learning models has been discussed. For example, \citeauthor{9020560}~\cite{9020560} compare the most common encoding schemes for predicting security events using logs. The result shows that it is considerably harmful to encode the representation without systematic justification. Similarly, \citeauthor{DBLP:journals/bmcbi/HeP16}~\cite{DBLP:journals/bmcbi/HeP16} study the effect of two encoding schemes for genetic trait prediction. A thorough analysis of the encoding schemes has been provided, together with which could be better under what cases. However, those findings cannot be directly applied in the context of software performance learning, due to two of its properties:

\begin{itemize}

\item Sampling from the configurable systems is rather expensive~\cite{DBLP:conf/icml/ZuluagaSKP13,DBLP:journals/corr/abs-1801-02175,DBLP:conf/mascots/JamshidiC16}, thus the sample size is often relatively smaller.
\item Software configuration is often sparse, i.e., the close configurations may have rather different performance~\cite{DBLP:journals/corr/abs-1801-02175,DBLP:conf/mascots/JamshidiC16}. This is because options like \texttt{cache}, when enabled, can create significant implications to the performance, but such a change is merely represented as a one-bit difference in the model. Therefore, the distribution of the data samples can be intrinsically different from the other domains.

\end{itemize}

Most importantly, this work provides an in-depth understanding of this topic for learning software performance, together with insights and suggestions under different circumstances.

\section{Conclusions}
\label{sec: conclusions}


This paper bridges a gap in the understating of encoding schemes for learning performance for highly configurable software. We do that by conducting a systematic empirical study, covering five systems, seven models, and three widely used encoding schemes, giving a total of 105 cases of investigation. In summary, we show that

\begin{displayquote}
\textit{Choosing the encoding scheme is non-trivial for performance learning and it can be rather expensive to do it using trial-and-error in a case-by-case manner.}
\end{displayquote}

Our findings provide actionable suggestions and ``rule-of-thumb'' when a thorough experimental comparison is not possible or desirable. Among these, the most important ones over all models and encoding schemes are:

\begin{itemize}
    \item using neural network paired with one-hot encoding for the best accuracy.
    \item using linear regression paired with scaled label encoding for the fastest training. 
    \item using scaled label encoding for a relatively well-balanced outcome, but mind the underlying model.
\end{itemize}

We hope that this work can serve as a good starting point to raise the awareness of the importance of choosing encoding schemes for performance learning, and the actionable suggestions are of usefulness to the practitioners in the field. More importantly, we seek to spark a dialog on a set of relevant future research directions for this regard. As such, the next stage on this research thread is vast, including designing specialized models paired with suitable encoding schemes or even investigating new, tailored encoding schemes derived from the findings in the paper.

\bibliographystyle{ACM-Reference-Format}
\bibliography{references}

\end{document}